\documentclass[12pt,preprint2]{aastex}        
 
\def\kms{\mbox{km/s}}  
\def\kpc{\mbox{kpc}}

\def\msun{\mbox{M$_\odot$}}  
        
\def\lsun{\mbox{L$_\odot$}}

\def\degree{\mbox{$^\circ$}}    
\def\mathnew{\mathsurround=0pt}    
\def\simov#1#2{\lower .5pt\vbox{\baselineskip0pt  
\lineskip-.5pt\ialign{$\mathnew#1\hfil##\hfil$\crcr#2\crcr\sim\crcr}}}     
\def\simgreat{\mathrel{\mathpalette\simov >}}    
   
\def\'#1{\ifx#1i{\accent"13\i}\else{\accent"13#1}\fi}    
  \def\et{et~al.}     
      
\begin{document}    
\slugcomment{{\em to be submitted to the Astrophysical Journal}}         
\shorttitle{Cores in Disk Galaxies: The Case of NGC~3109 and NGC~6822}     
\shortauthors{Valenzuela et al. 2006}

\title{Is there Evidence for Flat Cores in the Halos of Dwarf Galaxies?:    
\newline The Case of NGC~3109 and NGC~6822 }            

\author{Octavio Valenzuela  $^{1}$, George Rhee $^{2}$, Anatoly Klypin 
  $^{3}$, Fabio Governato$^{1}$, Gregory Stinson$^{1}$, Thomas 
  Quinn$^{1}$, James Wadsley$^{4}$ }

\affil{$^1$ Department of Astronomy, University of Washington  
Seattle, USA, WA 9819\\  
$^2$ University of Nevada, Department of Physics Box  454002, Las Vegas, NV 89154-4002 \\ 
$^3$ Astronomy Department, New Mexico State University, Box 30001, Department   
4500, Las Cruces, NM 88003-0001\\
$^4$ Department of Physics and Astronomy, McMaster University,
Hamilton, Ontario L88 2M1, Canada}

\begin{abstract}

Two well studied dwarf galaxies -- NGC~3109 and NGC~6822 -- present some
of the strongest observational support for a flat core at the center
of galactic dark matter (DM) halos.  We use detailed cosmologically
motivated numerical models to investigate the systematics and the
accuracy of recovering parameters of the galaxies.  Some of our models
match the observed structure of the two galaxies remarkably well.  Our
analysis shows that the rotation curves of these two galaxies are
instead quite compatible with their DM halos having steep cuspy
density profiles. The rotation curves in our models are measured using
standard observational techniques, projecting velocities along the
line of sight of an artificial observer and performing a tilted ring
analysis. The models reproduce the rotation curves of both galaxies,
the disk surface brightness profiles as well as the profile of 
isophotal ellipticity and position angle.  The models are centrally  
dominated by baryons; however, the dark matter component is globally  
dominant. The simulated disk mass is marginally consistent with a 
stellar mass-to-light ratio in agreement with the observed colors and  
the detected gaseous mass.  We show that non-circular motions combined    
with gas pressure support and projection effects results in a large   
underestimation of the circular velocity in the central   
$\sim 1$~kpc region, creating the illusion of a constant density core.    
Although the systematic effects mentioned above are stronger in barred systems, 
they are also present in axisymetric disks.   
Our results strongly suggest that there is no contradiction between    
the observed rotation curves in dwarf galaxies and the cuspy central    
dark matter density profiles predicted by Cold Dark Matter models.   
\end{abstract}


\section{Introduction}     
\label{sec:intro}

Cosmological models based on the inflationary paradigm, Cold Dark
Matter (CDM), and dark energy ($\Lambda$CDM) are very successful in
explaining the large scale structure of the Universe \citep{WMAP}.
However, it has been widely argued that $\Lambda$CDM models face a
number of challenges at galactic scales
\citep{Moore94,FP94,dBM97,McGaugh1998,Moore2001,Ostriker03}.  One of
the most persistent problems is the apparent incompatibility of galaxy
kinematics with the structure of dark matter halos predicted by
cosmology.  Cosmological N-body simulations predict halos with a
central cusp \citep{NFW97}. Because low surface brightness (LSB) and
dwarf galaxies are considered dark matter dominated at all radii, it
is expected that their dynamics provide an excellent benchmark for
testing the structure of CDM halos. It has been claimed that
cosmological halos disagree with the shallow dark matter density
profiles inferred from the rotation curves of dwarf and LSB galaxies
\citep{Moore94,FP94,Burkert95,dBM97}. The galaxies seem to favor dark
matter halos that have a soft core while cosmological halos are
cuspy. The analysis of the cusp-core controversy has motivated a large
amount of discussion and work
\citep[e.g.,][]{Firmani2000,AR01,Swaters03,swaters03b,
SalucciDDO47,Simon04,Rhee}.

Earlier, the beam smearing in radio observations was believed to be 
responsible for some discrepancy between the observations and the theory 
\citep{Bosch2000N3109,vandBosch01}. Later, high angular resolution  
observations \citep[e.g.,][]{dBBM2002, 
swaters03b,Weldrake2003,Simon04}, produced mixed results. In some  
cases rotation curves could be fitted with cuspy CDM halos, in others they 
could not.  \citet{swaters03b} concluded that in a statistical sense the 
observed population of LSB galaxies is consistent with cosmological 
predictions. Only some dwarf galaxies in their sample are clearly 
inconsistent with cuspy cosmological halos. \citet{hayashio4lsb} 
arrived at similar conclusions. \citet{BlaisN3109}  
presented mass models for a sample of dwarf galaxies. For example, the 
data for NGC 5585 are compatible with a cuspy dark matter 
halo. However, the halo concentration is small and the stellar 
mass-to-light ratio (M/L) is null. \citet{Swaters03} presented  
evidence for non-circular motions in the galaxy DDO 39. After applying 
simple corrections to the rotation curve, the galaxy appears   
compatible with a cuspy dark matter halo assuming a (M/L) equal to one,  
but still favoring a relatively low concentration. \footnote{Recent results from the 3rd year WMAP  
revision favor a smaller value of $\sigma_8=0.77$ making the concentration found by 
\citet{Swaters03} closer to $\Lambda$CDM predictions. See section~\ref{sec:discussion}. }   

 \citet{Rhee} showed that systematic  
effects related to disk thickness, bulges and bars can produce from 
mild  (10-20)$\%$ to  severe ( 30-50)$\%$ underestimation of the central  
circular velocity of galaxies, creating the illusion of constant density cores, 
while the actual  density distribution is cuspy.   Similar conclusions were reported by 
\citet{hayashi04}. In this case, the effect is associated with triaxial 
dark matter halos which produce disk ellipticity.       

 Recently, \citet{Simon04}, and \citet{Begum} reported results for low  
mass galaxies consistent with cuspy dark matter halos. In particular,  
\citet{Simon04} found evidence for significant radial motions or  
elliptical streaming in some galaxies. They argue that the  
non-circular motions can be related to halo ellipticity along the  
disk plane.  They also present one galaxy that favors a steep   
density profile assuming a stellar (M/L) consistent with the 
stellar population theory.  
\citet{Spekkens05} showed that systematic effects  
associated with long-slit observations are able to reproduce the             
distribution of slopes obtained in actual observations of low mass       
galaxies with no clear evidence for bars or bulges.  \citet{Dutton}      
discussed uncertainties in the stellar M/L ratio, halo triaxiality, and        
distance. They argued that those uncertainties may produce strong            
degeneracies, making estimates of halo central density less certain.           
\citet{Dutton} also presented estimates of the stellar M/L based on the          
B-R color and the stellar population models of \citet{Bell} and            
another estimate based on the maximum disk solution. \citet{Dutton}          
found that a disk contribution to the circular velocity based on the    
B-R color of NGC5585 is {\it above} the maximum disk solution.  This   
suggests that the gas in the center of this galaxy is not supported only by its     
rotation.  Note that internal reddening can produce similar results   
\citep{Dutton}.

 To summarize, uncertainties in measurements and interpretation of
 rotation curves can be quite substantial. In some cases cuspy dark
 matter profiles provide acceptable fits; in other cases the cusps are
 problematic.  Systematic errors are the most difficult part of the
 problem.  Models, which are disfavored by large $\chi^2$, can have
 circular velocities deviating by $\sim$ 10 $\%$ from the best-fit
 models. It is essential to carefully model and understand
 the systematics in order to decide if there is a conflict with
 cosmological predictions. In this paper we will focus on two types of
 systematic effects, that make the rotation velocity differ from the
 circular velocity: the pressure support and the elliptical motions.
 We also consider the projection effects
  discussed in \citet{Rhee}.

Because some galaxies are consistent with cusps, the heat of debate of cusps vs. cores shifts to 
those galaxies for which the discrepancy with cosmological predictions  is so large that it seems 
impossible to explain the observed kinematics with a 
cosmologically motivated halo even considering the uncertainties in the parameter  
estimation \citep[e.g.,][]{BlaisN3109,Moore2001,Weldrake2003,Simon04,Dutton}.  
The Magellanic-type galaxy NGC~3109 is the prototype  
of this kind of system. The galaxy is located at the outskirts of the 
Local Group.
NGC~6822 is another Local Group member of this type. Because of their 
nearby location, both galaxies have been subjects of high resolution studies,  
including their kinematics.  
 In order to illustrate the degree of disagreement with cosmology, we 
quote \citet{Weldrake2003}, who find that a model of NGC~6822 with a 
realistic DM halo gives an outstandingly bad fit: $\chi^2_{\rm reduced}=1200$.  
The goal of this paper is to design relatively 
detailed models for NGC 3109 and NGC 6822, and use the models to evaluate 
the severeness of the conflict that they pose for the $\Lambda$CDM model.

The standard way of constructing mass models for galaxies
 is to adjust parameters of analytical functions until the model
 circular velocity produces acceptable fits to the observed rotation
 curve.  However, this approach suffers from some shortcomings.  It is
 usually difficult to test the self-consistency of models: whether the
 observed gas and stellar kinematics are consistent with the assumed
 mass distribution. Disk stability against bar formation is another
 self-consistency test.  These potential caveats are known and have
 been used before to reduce the model parameter space in rotation
 curve analysis \citep{LiaMaxdisk,Fuchs}.  However, these problems are
 often ignored. It is not infrequent to see in literature that models
 that explain observed rotation curves have a maximum disk that is
 likely to be unstable for bar formation. In other situations,
 axisymmetric models are used to explain the kinematics of barred LSB
 galaxies, as pointed by \citet{swaters03b}.  In many of these cases,
 questionable results are taken as an evidence for cored halos.

Motivated by all these caveats and  
by the frequency of non-circular motions and non-axisymmetric structures in 
the observations of dwarf galaxies, 
in this paper we consider barred models, and study the corresponding non- circular motions. 
Given that many of the dwarf irregular galaxies show photometric or kinematic 
lopsidedness, we also consider this kind of models. In addition, we use the  
constraints on the stellar mass provided by observed colors of the 
stellar population.  We make high resolution simulations with several 
millions of particles moving in a self-consistent gravitational field 
for billions of years, that assure dynamical self-consistency. Once 
the simulations finish, we ``observe'' the models closely   
mimicking procedures applied to real observational data 
\citep{Rhee}. Models constructed in this way include a number of
effects, which are normally difficult to account for analytically: projection
effects, finite disk thickness, effects related to complex
non-circular motions, adjustment of the dark matter to the presence
and evolution of baryons. This procedure can be applied to pure
gravitational systems, which allow re-scaling and make it possible to fit
a specific galaxy.   

There are reasons to believe that the rotational velocities of the gas 
and  stars in dwarf galaxies are not very different. 
\citet{Rhee} and \citet{Hunter2002} addressed this issue
observationally.  They found that in the central $\sim 1$~kpc region
of late-type (nearly bulge-less) galaxies, stellar and gas rotation
curves are very similar to each other. At larger distances, gas
rotates slightly faster than the stellar component that, as one would
naively expect, has some intrinsic velocity dispersion. Judging by the
magnitude of the rotational velocity, one expects that in dwarf
galaxies such as NGC~3109 the  random velocities are small:
10-20~\kms, which is close to the random velocities of HI 
clouds. Indeed, this is what we find in our simulations.
Hence, it is reasonable to use collisionless models. 
However, in order to validate our methodology, 
we also performed state of the art $N-$body and hydrodynamical 
simulations. The simulations included a realistic implementation of 
stellar and supernova (SN) feedback and allow us to study the effects of gas 
motion in dwarf galaxies.

The paper is organized as follows.  In Section~\ref{sec:observations} 
we review the observational properties of NGC~3109\ and NGC~6822. In
Section ~\ref{sec:hydro} we present our hydrodynamical models of a
dwarf galaxy and compare the gaseous and the stellar kinematics. 
Models for NGC~3109 and NGC~6822 are
presented in Section~\ref{sec:Models}. A discussion of our results is
given in Section~\ref{sec:discussion}. A short summary of our results
is presented in Section~\ref{sec:conclusions}. Appendix gives more
details of gas motion in hydrodynamical models. We also discuss the
corrections to the gas rotation velocity necessary to recover the
circular velocity.

\begin{deluxetable}{lcccc}    
\tablecolumns{5}  
\tablewidth{0pt} 
\tablecaption{Observed Properties of NGC~3109 and NGC~6822}  
\tablehead{\colhead{Parameter} & \colhead{NGC~3109}\hfill & Reference 
                               &\colhead{NGC~6822}\hfill & Reference \\ } 
\startdata  
Morphological type              &   SB(s)m        & 1    &   IB(s)m & 1\\    
Distance                        & 1.36~Mpc        & 2    &   0.49~Mpc & 8,9,10     \\    	       
Inclination                     & $75\degree\pm 2^o$   & 3    &   $60\degree\pm 2\degree$   &  8,11 \\
B-band magnitude                & -16.35          & 4    &   -15.8    & 11  \\     
K-band magnitude                &                 &      & -17.9      & 8\\  
B-R color                       &  $0.8\pm 0.16$  & 5    &  0.8       & 8 \\     
Exponential disk scale-length $r_{\rm d}$ 
                                & 1.2~kpc         & 4    &  0.68~kpc  & 8 \\
Maximum rotational velocity     & 67~\kms         & 3,6  &  55~\kms   & 8 \\      
Mass of neutral hydrogen M$_{HI}, \msun$ 
                                & $(4.9\pm 0.6)\times 10^8$ & 7 
                                & $1.34\times10^8$ &     12,14    \\         
Mass of gas, \msun
                                & $(10.3\pm 0.9)\times 10^8$ & 7,13 
                                & $2.3\times10^8$ &     12,14,15    \\

\label{tab:observations}              
\enddata 
\tablerefs{(1)~\citet{deVaucouleurs91} (2)~\citet{Musella1997} (3)~\citet{Jobin} (4)~\citet{CarignanN3109} (5)~\citet{colors} 
            (6)~\citet{BlaisN3109} 
            (7)~\citet{Barnes01} (8)~\citet{Weldrake2003} 
            (9)~\citet{Mateo98} 
            (10)~\citet{Gottesman77} (11)~\citet{1991PHodge} (12)~\citet{deBlokWalter} 
            (13)~\citet{Kahabka00}   
            (14)~\citet{dBW2006} (15)~\citet{Israel1997}  
                       }
\tablecomments{{\it NGC~3109:} Results were rescaled assuming the distance of 1.36~Mpc;
absolute magnitude was corrected for internal absorption \citep{CarignanN3109}.
              }
\end{deluxetable}

\section{Observational Data}    
\label{sec:observations}

\subsection{NGC~3109}     

NGC~3109 is a Local Group member that has been extensively studied
since it is one of the nearest dwarf irregular galaxies.
Observational data for the galaxy are presented in
Table~\ref{tab:observations}. Mass of gas in NGC~3109 is an important
parameter for our models. \citet{Barnes01} estimated the neutral
hydrogen mass for NGC~3109 to be $3.8\times 10^8\msun$ (after
correction to the distance to NGC~3109).  Mass of
molecular gas is more uncertain.  \citet{Kahabka00} found that not more
than 60\% of hydrogen in NGC 3109 can be in a molecular phase.  Based
on CO observations \citet{Rowan-Robinson80} find that more than 10\%
of HI must be in molecular phase.  If we take that 30\% of the total hydrogen is
molecular, which is reasonable following \citet{Leroy2005}, 
and use a factor  1.4 to account for helium and metals, 
the gas mass in NGC~3109 is estimated to be $(10.3\pm 0.9)\times 10^{8}\msun$. 

 \citet{Jobin} analyzed observations in I-band as well as in radio
wavelengths. The I-band photometry shows a change in the position
angle and ellipticity of isophotes as a function of radius, suggesting
the presence of a stellar bar, in agreement with B-band observations
\citep{CarignanN3109}.  \citet{Jobin} concluded that the system is
globally dominated by dark matter and their favored model requires a
halo with a core radius of almost 7~kpc. \citet{Bosch2000N3109} fit an NFW model with a
central constant density core assuming the stellar $(M/L)=0$. 
 Finally, \citet{BlaisN3109} analyzed the two-dimensional
kinematics of NGC~3109 based on high resolution H$_\alpha$
observations. Their best model with a cuspy halo over-predicts the
rotation in the central kiloparsec. This model requires a negligible
stellar (M/L) ratio and a concentration smaller than one, which is
difficult to reconcile with cosmological predictions.  Their favored
model has a constant density core radius extending up to 2.4~kpc. 
Interestingly, the size the halo core has decreased  as the 
observations and the analysis have been improved.  Given the quality of 
the observational data in \citet{BlaisN3109}, it is considered that 
NGC~3109 presents a serious problem for cosmological models 
\citep{SalucciDDO47}.
 
\subsection{NGC~6822}              
NGC~6822 is also a Local Group member and the third nearest dwarf
irregular galaxy after the LMC and SMC. Extensive analysis of properties of the
ISM and the stellar component for this galaxy has been done by
\citet{deBlokWalter,Weldrake2003,Komiyama03,dBW2006}.  In order
to estimate the total gas mass, we use \citet{Israel1997} results on
the mass of molecular gas $M(H_2)=0.4\times 10^8\msun$ and make a
correction for helium and metals.
A summary of the observational properties used in this paper is presented in
Table~\ref{tab:observations}.

 Though the galaxy is classified as barred, it is generally considered
 as ``a rather average and quiescent dwarf irregular galaxy''
 \citep{dBW2006}. This probably explains why \citet{Weldrake2003} used 
 simple models (no bar or any other complications) to fit the HI
 rotation curve of this galaxy.  Their best model with a cuspy dark
 matter halo assumes a zero stellar mass-to-light ratio and a
 concentration close to zero, which is hard to justify not to mention
 that it systematically over predicts the rotation velocity inside the
 central kiloparsec. A model with a reasonable halo concentration was
 rejected with very high confidence.

Yet, the galaxy is far from being simple.  The stellar distribution
 shows a short, but rather strong bar, which is prominent in visible
 light. This is clearly seen in Figure~8 of \citet{dBW2006}:
 the distribution of light in the central $10' \approx 1.5~\kpc$\ is
 misaligned with the HI at large radii. The misalignment is quite
 large: \citet{1977PHodge} and \citet{Cioni} give the position angle
 of the bar $PA=10\degree$, while the position angle of the HI disk is
 about $PA=125\degree$ \citep{Weldrake2003}. This bar was considered
 the main stellar body of the galaxy; it has an exponential
 profile. However, deeper photometry reveals a fainter blue component
 tracing the HI distribution \citep{Komiyama03}. There is also a faint
 and rounder red stellar component extending to several radial scale
 lengths \citep{Letarte02}. The neutral hydrogen and the faint blue
 stars show a different position angle compared with the stellar bar.
 The central surface brightness in B-band is 27~mag$/{\rm
 arcsec}^2$. This is comparable to the surface brightness of large LSB
 galaxies \citep{Komiyama03}. All these properties make NGC~6822 one
 of the nearest barred LSB galaxies. However, the bar has not been
 considered in previous analyses of NGC~6822 kinematics
 \citep{Weldrake2003}.

\clearpage

\begin{figure*} 
\epsscale{2.3} 
\plottwo{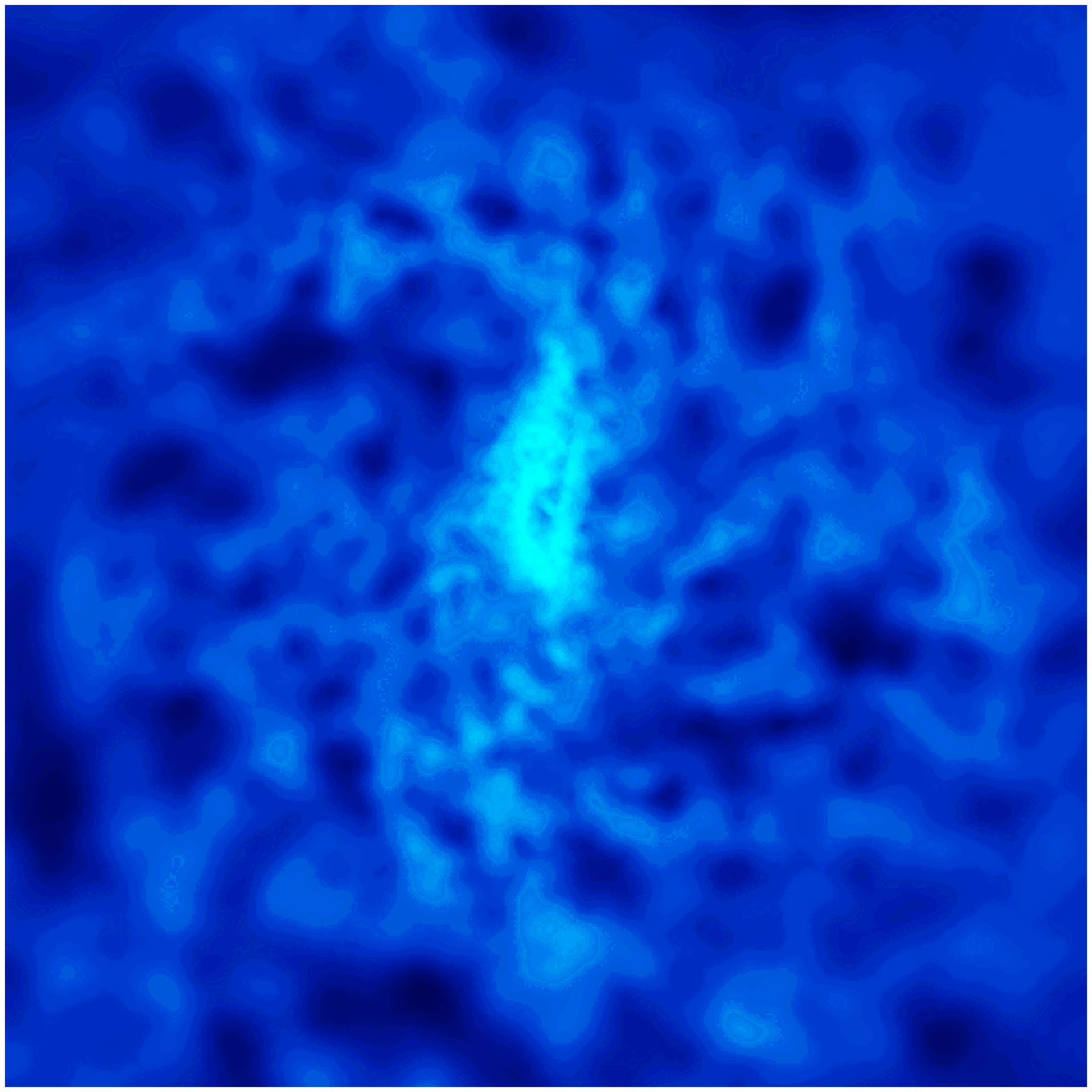}{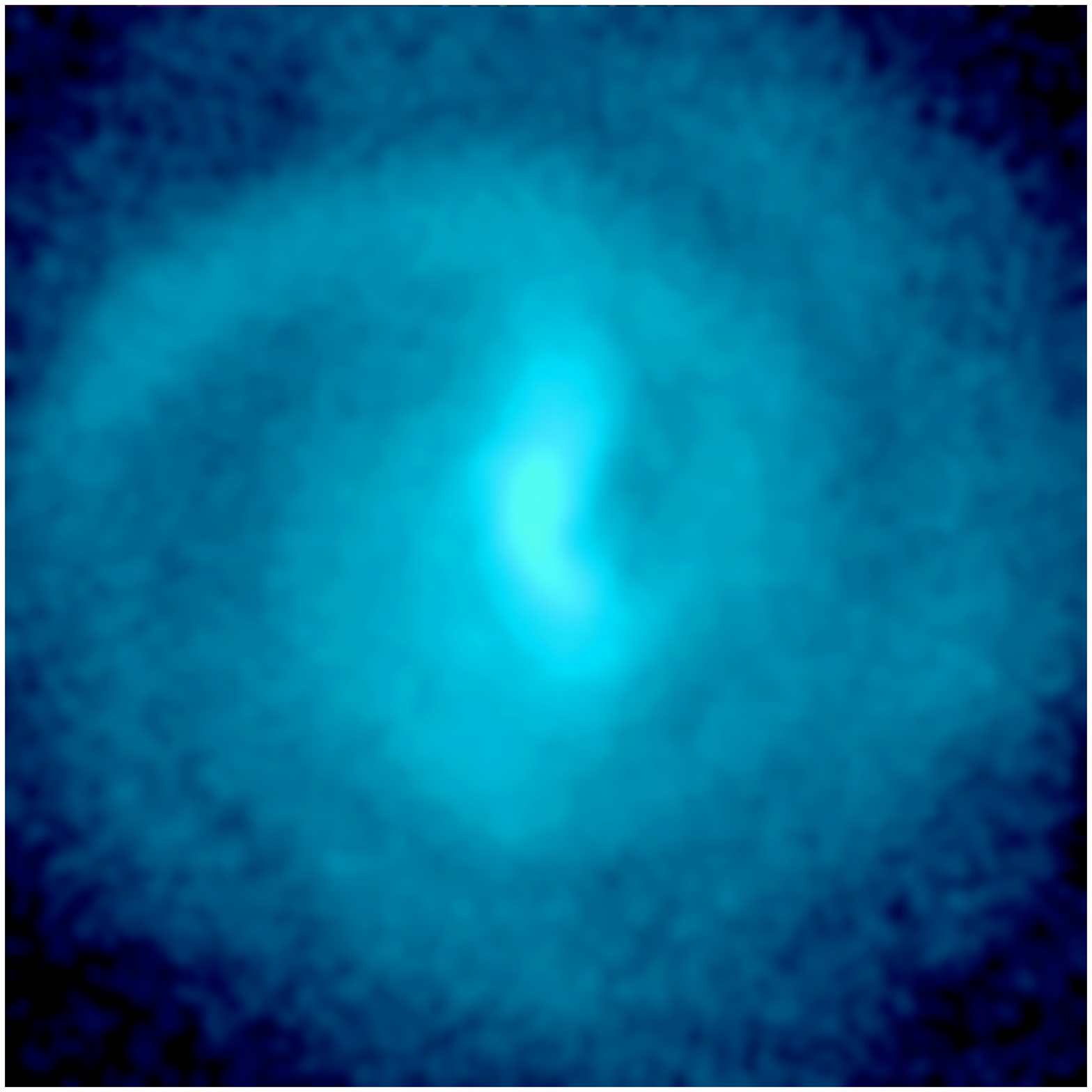}  
\epsscale{2.7} 
\plottwo {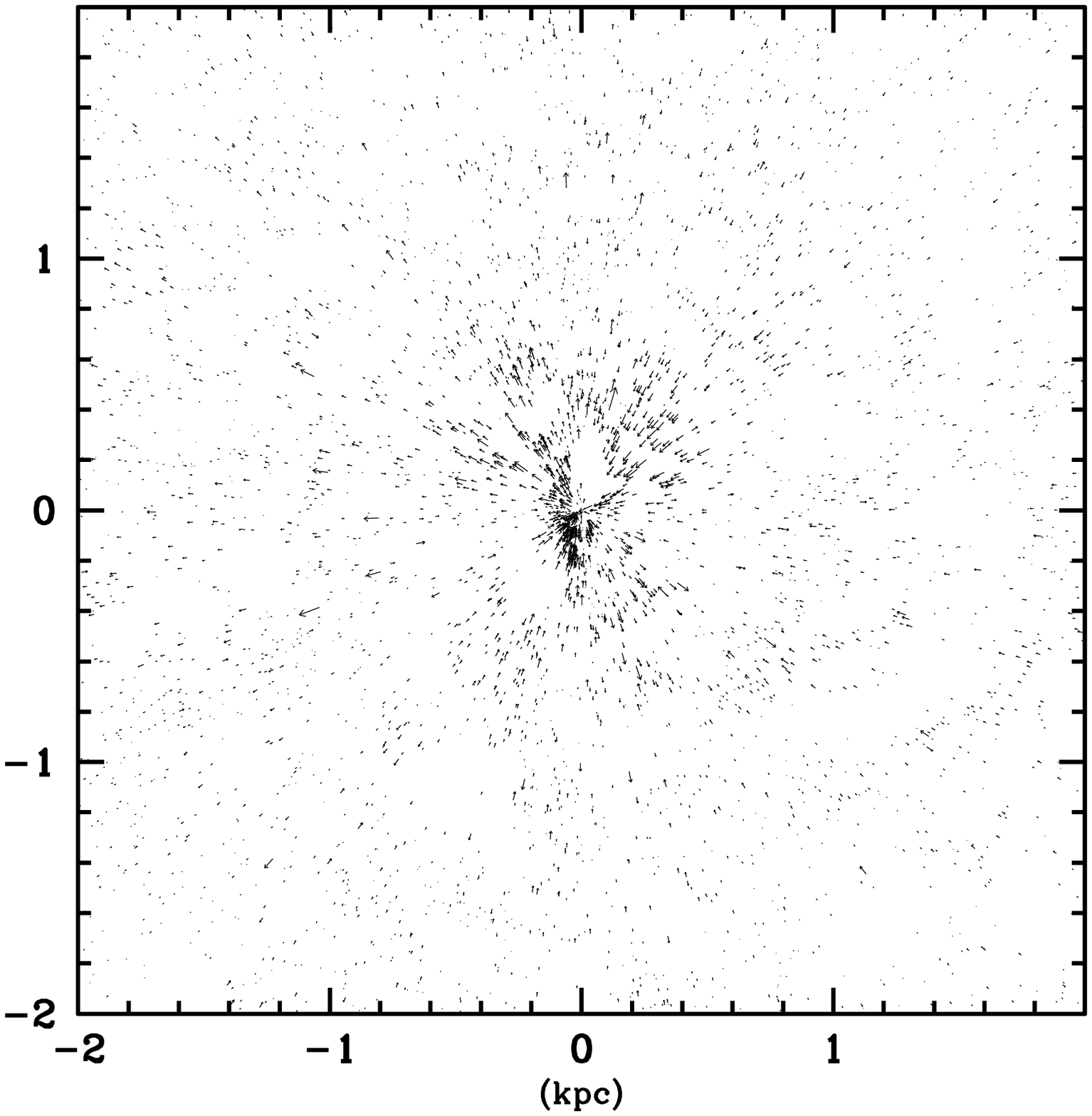}{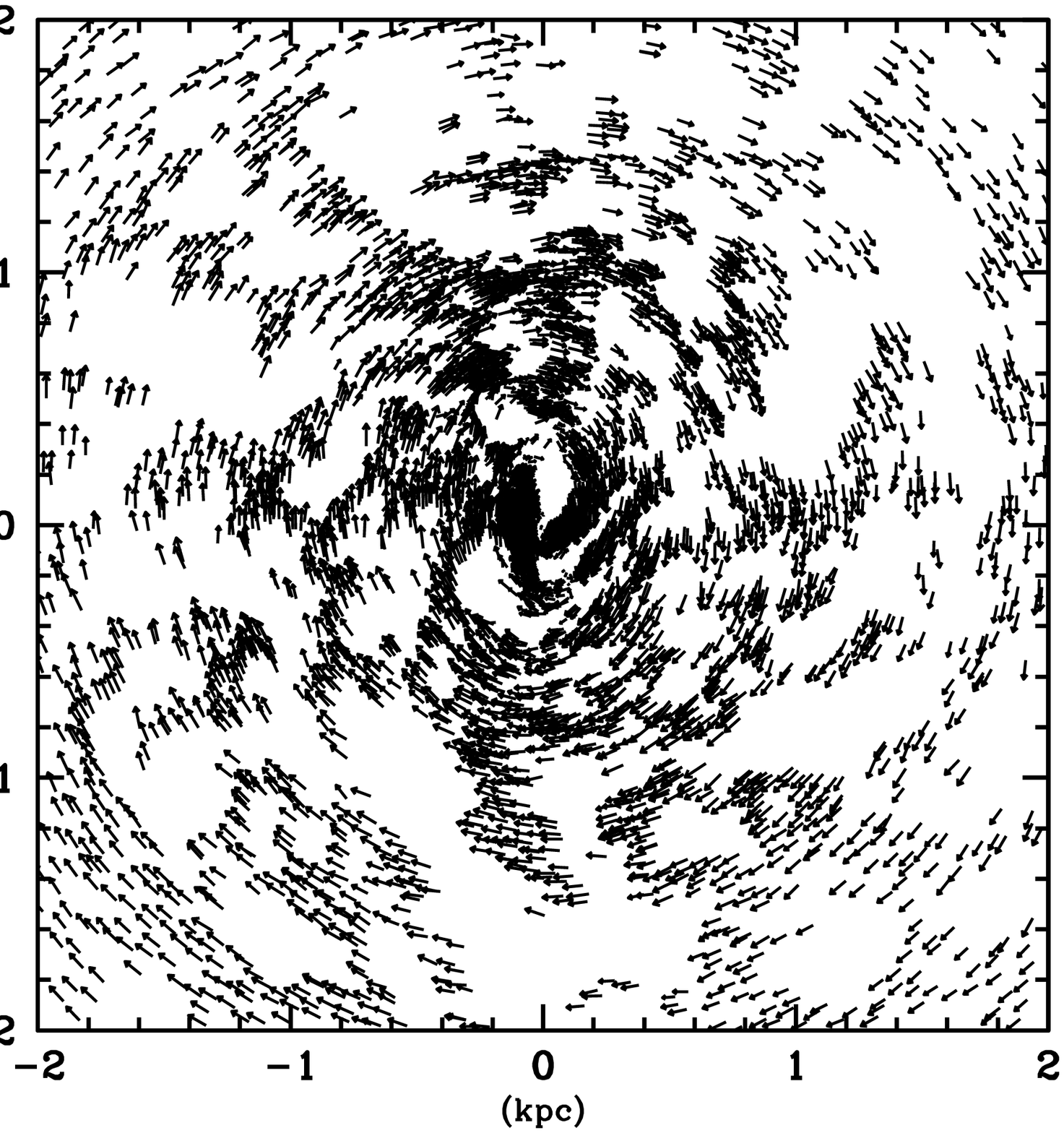} 
\caption{The structure and kinematics of the disk 
in the hydrodynamic simulation. The upper left panel shows the face-on view of
cold gas with $T\leq 15000$~K. The cavities are full with 
a warm gas component that is not shown in the figure ($T\approx (2-10)\times 10^4$~K).
The upper right panel shows the distribution of the stellar component. Notice that the bar in the 
gaseous component is weaker and shorter than in the stellar component.  
The lower left panel shows the radial velocities of cold gas and the lower 
right panel shows the full velocity field. An arrow with a 
length of 100~pc corresponds to velocity 60~km/s. The systematic 
radial motions are clearly present, but their amplitude is low 
compared to the 42~km/s rotation velocity of the gas at 500~pc and 
with the 20~km/s value of the rms velocities at the same radius.  
The upper panels show the projected density along the line of sight. 
Lower panels present particles in a slice of 20~pc thickness.}

\label{fig:hydro}
\end{figure*} 

\clearpage

\section{Kinematics of Gas and Stars: Hydrodynamical simulations.}  
\label{sec:hydro}            
In order to test the validity of our conclusions based on  
collisionless simulations (see also \citet{Rhee}), we made high  
resolution hydrodynamical simulations of a dwarf galaxy.  These  
simulations cannot be rescaled because they include gas dynamics and   
radiative processes.  Therefore, fitting them to a specific galaxy is  
difficult. However, they can shed light on the physical processes  
playing a role in dwarf galaxies with a significant gas component.  In  
this section we are mostly interested in a simulation, which produced 
a weak bar because this comes along with what we find in NGC~3109 and 
NGC~6822. Results of other simulations are presented in the APPENDIX.

The models include a dark matter halo initially with the NFW density
profile modified by the standard adiabatic contraction 
\citep{Blumenthal86}. The initial parameters of the simulations are 
presented in Table~\ref{tab:hydro}. The models are set initially in 
equilibrium. The choice of parameters for the simulations was 
motivated by parameters of NGC~3109 and NGC~6822.  The main parameter 
is the maximum circular velocity. After evolving for 1~Gyr the model 
had a maximum circular velocity 70~\kms, which is very close to that 
of the real galaxies. Just as in real galaxies, the disk-to-halo ratio in 
our models is very small.  In next sections we show that our best 
models for the galaxies have $M_{\rm disk}/M_{\rm vir}\approx 
0.025$. We selected this value for our hydrodynamical models. Note 
that this is almost 7 times smaller than the average ratio of baryons 
to dark matter in a $\Lambda$CDM universe.  Even with this small mass 
the disk makes a significant contribution to the circular velocity in 
the central 1/3~kpc. Though a reasonable match to real galaxies for 
some parameters, the hydrodynamical model does not reproduce all the properties of 
dwarf galaxies. Most notably, the disk is more compact than the disk 
of the galaxies that we are studying.

The simulations were evolved with the multi-step parallel Tree/SPH code 
GASOLINE \citep{gasoline}.  The code includes hydrodynamics, star 
formation, radiative and Compton cooling for a primordial mixture of 
hydrogen and helium as well as stellar and SN feedback.  Feedback is 
modeled as thermal energy dumped into the gas near to star 
particles. The deposition rate is tied to stellar lifetimes.  The 
affected gas has its cooling shut off for 20~million years, mimicking 
pressure support from (unresolved) internal turbulence 
\citep{thacker00}. The only free parameters (star formation efficiency  
and the fraction of SN energy dumped into the interstellar medium 
(ISM) have been tuned over a range of galaxy masses to reproduce a 
number of observables \citep{Governato2006,Stinson 2006}. 
Those include the shape and normalization of the Schmidt law of  
our own Milky Way galaxy and the typical disk thickness, the star  
formation rates and the gas turbulence observed in small galaxies. 
Our simulation has initially 2~million particles and has 30~pc softening  
for gas and stellar component and 60 pc for the dark matter particles. 
The force is Newtonian at $\sim$ 60 pc. We follow the evolution 
only for 1~Gyr.  Dynamically, this is a very long period  for the 
dwarf ``galaxy'': 11 orbital periods at 1~kpc distance.

 During the period of bar formation the halo of model H1 is
contracted adiabatically due to the disk evolution
\citep{Colin06}.  The bar forms very quickly (in about 100~Mrs).  
After that initial period the system is in a quasi-stationary state: Gas show 
only small radial velocities (see figure~\ref{fig:hydro}). 
Although we analyze the simulation at $\sim$ 1 Gyr there is
not a particular reason for that.  
The model H2 was simulated twice: with and without the star formation. 
The disk in this model is more extended than in the model H1. As the result, the model
is  dominated by the dark matter even in the central
region. The model did not produce a bar after 1~Gyr of evolution.

\begin{figure*}[tb!]   
\epsscale{2.2}  
\plottwo{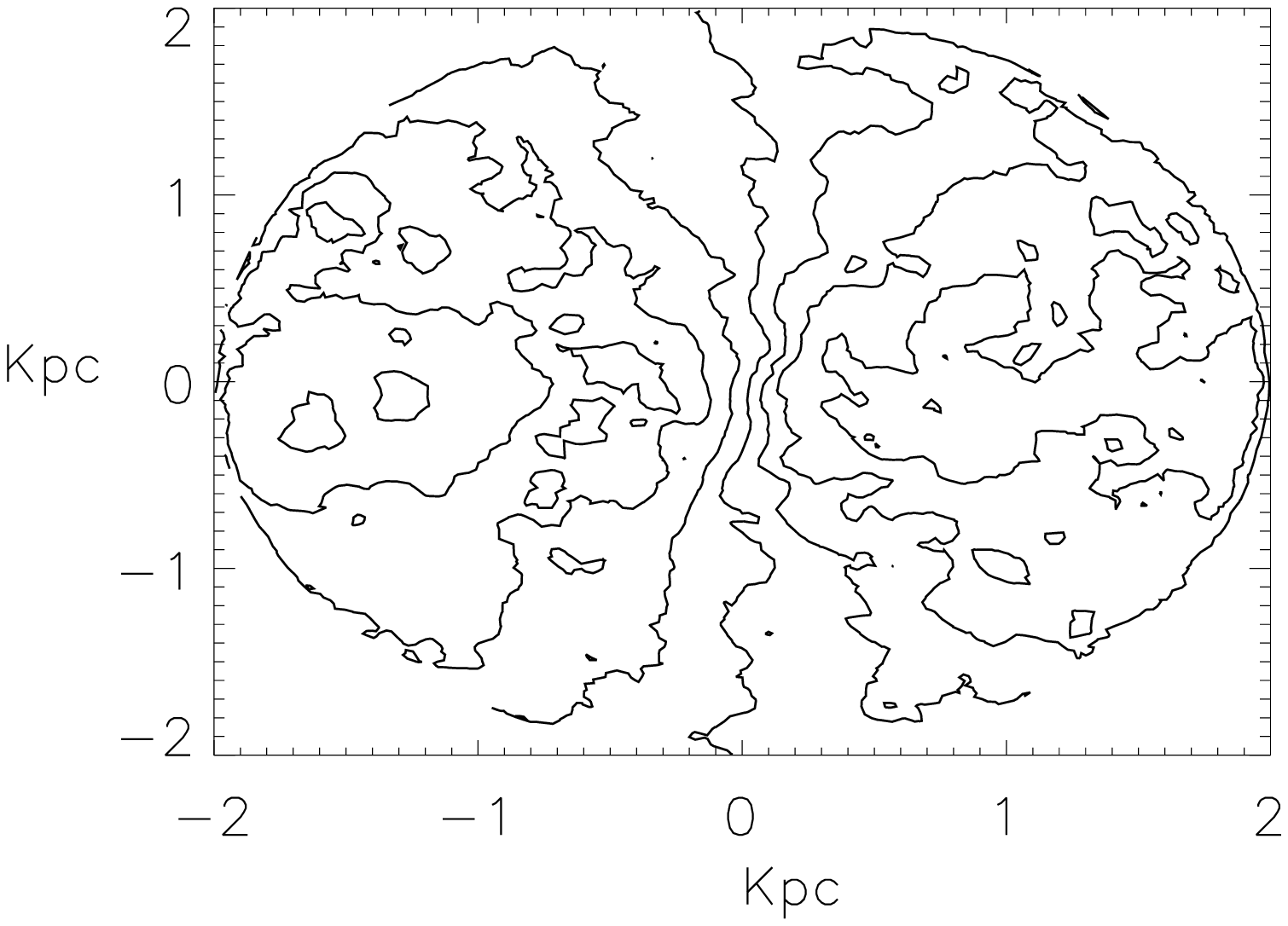} {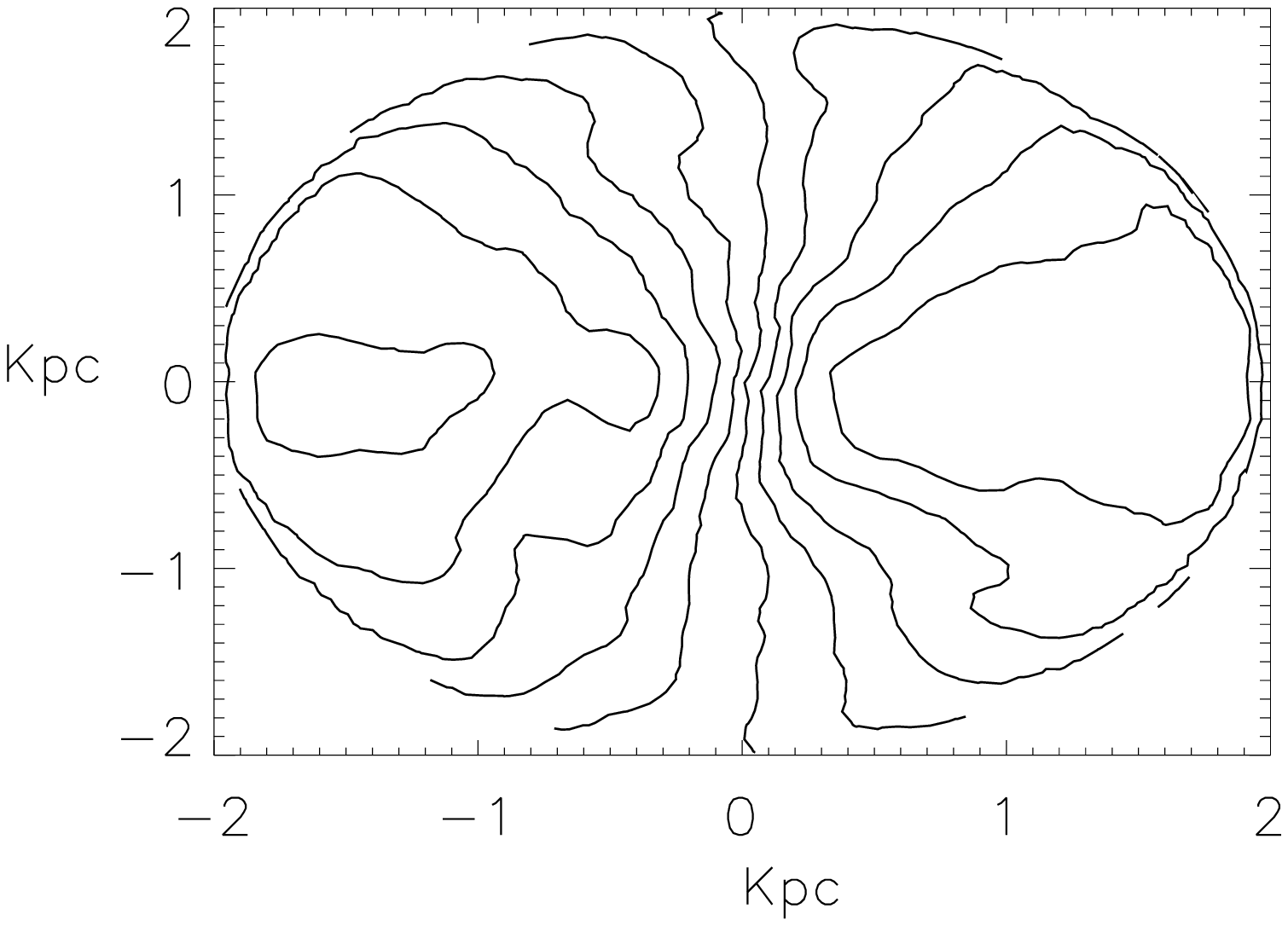}      
\caption{The two-dimensional velocity field of cold ($T< 15,000$~K)    
gas in the hydrodynamic model H1.  The figure shows the contours of  
constant velocity along the line of sight for an observer that  
measures an inclination of $60\degree$ for the galaxy. The left and   
right panels use different smoothing of the velocity field. The noisy   
behavior of contours in the left panel is due to super-bubbles, which  
induce $5-10\kms$ motions in the surrounding cold gas. The smoothed  
velocity field in the right panel shows that kinematic major and minor  
axis are not orthogonal, which is due to oval distortions related with  
$\sim 1$~kpc bar. The bar has a position angle of $\sim 15\degree$   
measured clock-wise from the vertical direction. }    
\label{fig:hydro-vfield}                
\end{figure*}

Figure~\ref{fig:hydro} shows the structure of the disk as
well as the velocity field and radial motions of the cold gas.  The
distribution of the cold gas is remarkably complex.  There is a weak
bar, which extends in radius to only 1~kpc. There are spiral arms.  At
the resolved scales the gas shows a complex structure with cold
filaments and holes filled with dilute and relatively hot gas.  At
these temperatures ($\sim 10^5$K) and densities the hot gas cools
relatively quickly. Consequently, there should be a source of energy and a
mechanism to replenish the hot gas. Both are provided by the star
formation.  The star formation rate $5.5\times 10^{-2}\msun/$yr is
relatively low and is comparable with observational estimates of
$(3-6)\times 10^{-2}\msun/$yr for NGC~6822 \citep{Wyder01}.

Large (100-300~pc) holes and lumps in the distribution of the cold gas 
observed in Figure~\ref{fig:hydro} are typical for small real 
galaxies.  For example,  Figure~8 in \citet{Weldrake2003}  
shows numerous 1-3~arcmin (100-300~pc) holes and lumps in the 
distribution of the neutral hydrogen for NGC~6822. \citet{Walter2001} give a 
catalog of 19 holes with size larger than 200~pc in another dwarf 
galaxy DDO~47. The same situation -- numerous holes, filaments, and 
lumps -- is found in the SMC \citep{Staveley1997,Hatzidimitriou2005}, LMC 
\citep{Kim1998}, Ho II \citep{Stewart2000}. In other words, the 
multi-phase medium, which we find in our simulation, is characteristic 
for dwarf irregular galaxies such as NGC~3109 and NGC~6822 studied in 
this paper.

The observed velocity field in the central $\sim~\!2~\kpc$ region of
model is presented in Figure~\ref{fig:hydro-vfield}.  The noisy
behavior of contours in the left panel is due to super-bubbles, which
induce $5-10~\kms$ motions in the surrounding cold gas. The same kind
of ``noise'' is observed in real galaxies, if observations are done
with high resolution. Only after significant smoothing we find a
normal ``spider diagram'' similar to Figure~1 in \citet{Swaters03}. In
this diagram the presence of the bar is a rather subtle effect -- a
twist of contours in the central 1~kpc area -- and can easily be
overlooked.  Qualitatively the same twist of contours in the velocity
field is observed in the NGC~6822 velocity map \citep[][Figure
6]{Weldrake2003}.  In order to quantify the magnitude of the motions,
we estimate the rms velocities in different parts of the model.  To 
some degree we follow the observational procedure. In observations the
velocity dispersion is often estimated using the gaussian width of 
emission lines.  This is done with some spatial filtering.  In order 
to mimic the observations, we estimate the one-dimensional velocity 
dispersion of the cold gas ($T<15,000~$K). We use different filter 
sizes (box sizes of 50, 100, and 200 parsecs) to make maps of the  
observed rms velocity across the whole galaxy. For all 
cases the average and the most frequent value for the rms velocity in
the pixels are 8 and 6~km/s respectively. Thus, local estimates of the 
velocity dispersion in our model are relatively small and are constant.  
This is compatible with what is measured in real galaxies at the scale of 
tens and few hundreds of parsecs \citep[e.g.,][]{deBlokWalter}. 
These values of the rms velocity would normally imply that gas motions 
are negligible and gas motion faithfully tracks the mass distribution.

\begin{figure}[tb!]   
\epsscale{1.1}     
\plotone{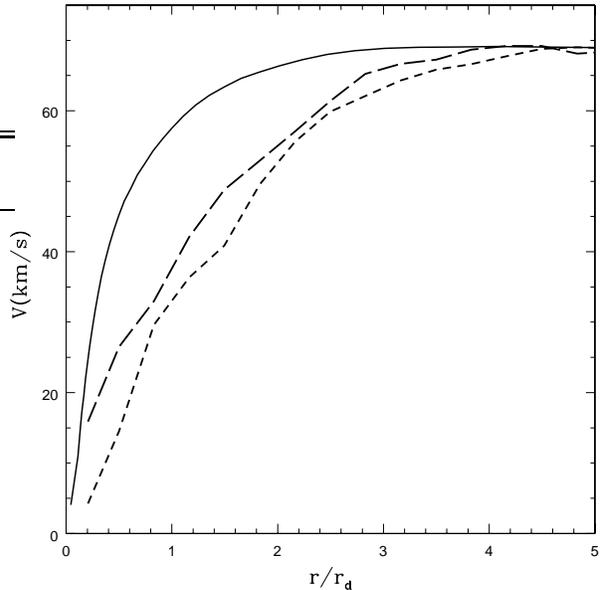}          
\caption{Velocity curves in the hydrodynamical simulation H1 of a dwarf  
galaxy.  The full curve shows the spherical averaged 
circular velocity $\sqrt{GM_{\rm total}/r}$. The long-dashed curve is for the 
azimuthally averaged rotation velocity of gas colder than 15000~K. 
The stellar rotation velocity is shown by the short-dash curve.  In
the central 1~kpc region the stars rotate slightly slower, but close
to the cold gas. Curves for gas and stars are substantially below the
circular velocity. At $r>3r_d$~kpc all the curves are practically the   
same.}  
\label{fig:hydrorotcur}  
\end{figure}

 Figure~\ref{fig:hydrorotcur} shows that this is not true: cold gas 
rotates slower than expected. The figure presents azimuthally 
averaged velocities of different components in the model H1. 
We show the ``true rotation curves'' \citep{Rhee}: velocities  
are defined for components in the disk plane. An observer would need   
to consider in addition projection effects that make the
rotation even smaller for most view angles \citep{Rhee}
(see section \ref{sec:Models} for details).  
We note that the disk is now a dominant component in the central 
region of the ``galaxy''.  The plot shows another important property:  
the cold gas and the stellar component rotate with similar speeds in the central 
region, in agreement with the observations reported by \citet{Rhee} 
and \citet{Hunter2002}.  The main difference between both curves are
features related with gas shock waves and an offset of few kilometers
per second in amplitude.

It is interesting to apply traditional estimates of the density 
reconstruction using the rotation curve of the cold gas of the Model H1. 
We use the following relation \citep{deBlok2001}
to find the total density $\rho(r)$ when the cold gas moves with velocity $V(r)$:
\begin{equation}
  4\pi G\rho(r) = 2\frac{V}{r}\frac{\partial V}{\partial r} +
  \left(\frac{V}{r}\right)^2.  \label{eq:rho} \end{equation} 

This equation assumes a neglegible disk, which is considered as reasonable 
for LSB galaxies. If the model were 
treated as a real galaxy, the non-circular motions would be estimated  
as negligibly small. We already mentioned the rms velocities $\approx 8~\kms$
estimated using  line widths. The average point-by-point variations 
of velocities also appear to be small: $\approx 10-12~\kms$.
Adding in quadratures any of those estimates to the ``observed''
rotation velocity makes very small effect on the rotation curve.  For
example, at $r=r_d$ the rotational velocity is $\approx
35~\kms$. Adding in quadratures $15~\kms$ gives $38~\kms$, which is
significantly smaller than the circular velocity of $\approx 60~\kms$
at that radius. We emphasize that the non-circular velocities only
appear to be small. This is how they would be measured by an
``observer''. In reality the magnitude of the non-circular motions is
larger and their effect is significant. APPENDIX gives more extended discussion of effects non-circular
motions and pressure gradients, as well as the corrections necessary to apply
to gas kinematics in order to recover the circular velocity.

\begin{figure}[tb!]  \epsscale{1.1} \plotone{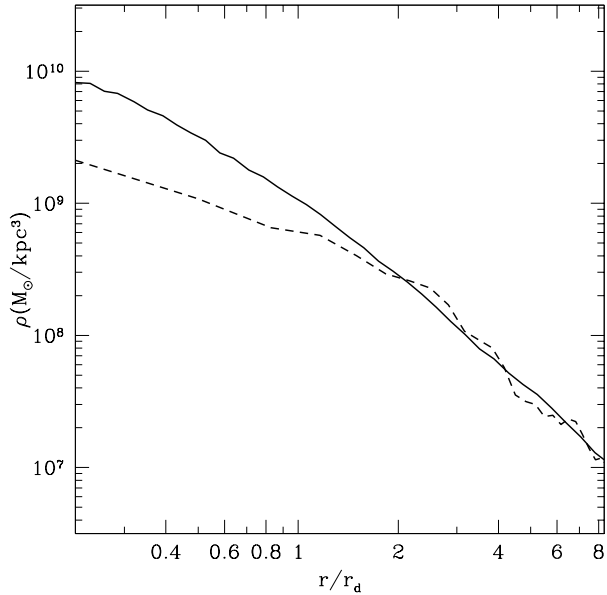} 
\caption{Spherically averaged total density profiles for the
hydrodynamical simulation H1 of a dwarf galaxy.  The full curve shows
the true density profile. The dashed curve is the recovered density,
calculated from  the true rotation curve of cold 
gas. At large distances $r>3r_d$ the rotation of the cold gas gives a 
good estimate for the true density in the system. Yet, at  
smaller distances the cold gas systematically and substantially 
underestimates the density. Different effects contribute to the 
mismatch.}  \label{fig:hydrodens} \end{figure}

Figure~\ref{fig:hydrodens} presents the true density profile and the 
recovered profile.  At distances $r>3r_d$ the rotation of the cold gas 
gives a good estimate for the true density in the 
system. Yet, at smaller radii the cold gas systematically and 
substantially underestimates the density. We emphasize that 
projection effects can make the density underestimation even 
stronger \citep{Rhee}.   

In the APPENDIX we give 
analysis of the discrepancy. We show that the true density can be 
recovered, but this requires accurate and careful account of numerous 
effects. Without this, one can get a wrong conclusion that the density 
in the central region has a core while it actually has a  
cusp. We also present results of two other simulations (two variants of 
the Model H2), which address issues of the resolution and the 
star-formation feedback. In the simulation without the stellar 
feedback the cold gas rotates very close to the circular velocity of  
the system. Thus, feedback is definitely responsible for a 
large fraction of the discrepancy. In the
simulation with feedback, but without a bar, the discrepancy is
present. Yet, it is not as strong as in the model H1, which has both
the feedback and the bar.

We draw two important conclusions from the analysis of the 
hydrodynamical simulations: (i) the cold gas can rotate slower than the
circular velocity. This happens in the central regions of galaxies due
to the combined effect of the stellar feedback and a weak bar.  (ii)
In the central regions the gas rotation and the rotation of the stars
are quite comparable. Thus, the stellar motions can be used as a proxy
for the cold gas rotation in N-body models.

\begin{table}[tb] 
\caption{Parameters of Hydrodynamical Simulations}
\begin{center}
\small
\begin{tabular}{lll}
\tableline\tableline\\
Parameter  & Model H1 &  Model H2 \\  
\tableline \\  
Virial mass                     $M_{200}$, ($\msun$)                   & $2\times 10^{10}$ & $3.4\times 10^{10}$ \\  
Virial velocity                 $V_{200}$                   & 40 \kms     & 40 \kms    \\ 
Number of dark matter                & $1.7\times 10^6$ & $2.7\times 10^6$  \\     
\qquad particles              \\     
Number of stellar                  & $2\times 10^5$  & $2\times 10^5$ \\   
\qquad particles   $N_{\rm disk}$           \\     
Number of gas                       & $1\times 10^5$ & $2\times 10^5$  \\   
\qquad particles         $N_{\rm gas}$        \\   
Disk to halo mass ratio          & 0.025   & 0.02 \\       
Halo concentration              $C_{\rm vir}$               & 10. & 10. \\         
Disk height scale               $z_{\rm d}$                       & 60 pc & 60 pc  \\        
Disk exponential scale                        & 300 pc & 600 pc \\  
 \qquad length   $r_{\rm d}$                  \\  
Disk stability parameter        $Q$                         & 1.2 & 1.2 \\     
Maximum Resolution    & 60 pc & 60 pc  \\       
  \qquad ($2\times\epsilon_{star}$)  \\        

\tableline
\label{tab:hydro}
\end{tabular}
\end{center}
\end{table}

\section{Models for Dwarf Galaxies NGC~3109 and NGC~6822}     
\label{sec:Models}   

\subsection{N-body Simulations}      
As we discussed in the introduction, we use  N-body simulations in 
order to model the mass distribution in NGC~3109 and NGC~6822. Our  
simulations were performed using the parallel Adaptive 
Refinement Tree (ART) code \citep{KKK97}. In this models the disk component is simulated using equal-mas stellar particles.
The dark matter halo is sampled with multiple mass species. The halo region surrounding the
disk and extending to $\sim$ three times the disk size is sampled with equal-mass particles, each particle having the same mass as
that of the ``stars''. At larger distances the particle mass increases  with radius 
as described in \citet{barpaper}: each subsequent species is twice as massive  as the previous one.
We use a total of five different mass species for model I and seven  mass species for model II. 
 Model~I was evolved for 8~Gyrs and model~II fas evolved for 3.4~Gyrs.   
Initially the models have only two components: an exponential disk and a  
dark matter halo with the NFW density profile.  The prescription for 
setting up the initial conditions  is described  in \citet{barpaper}.  The initial   
parameters of the models and the simulations are shown in Table~\ref{tab:nbody}.
Model~I  is based on the model A1 presented by \citet{barpaper}, model~II was
 designed from the beginning as a dwarf galaxy.  

In addition, the model~II was also run to produce lopsided models.
There are some motivations to study lopsided models.  NGC~6822 shows a
tidal tail-like features at large radii. NGC~3109 is also lopsided and
probably warped.  These features may produce additional non-circular
motions. We do not try neither to mimic the magnitude of observed  
lopsidedness nor to investigate the origin of it. We only want to find 
the effect of the lopsidedness on measured rotation curve.  In order to 
trigger the lopsidedness, we follow the prescription of \citet{Levine98}: 
We displace the disk from the center of the dark matter halo by 0.8 
and 2.5 times the initial disk scale length. We then follow the 
evolution of the system for 3~Gyrs. We call the model with smaller 
initial displacement ``mildly lopsided'' and the model with larger 
displacement ``strongly lopsided''.

\begin{figure*}[tb!] 
\epsscale{1.3} 
\plotone{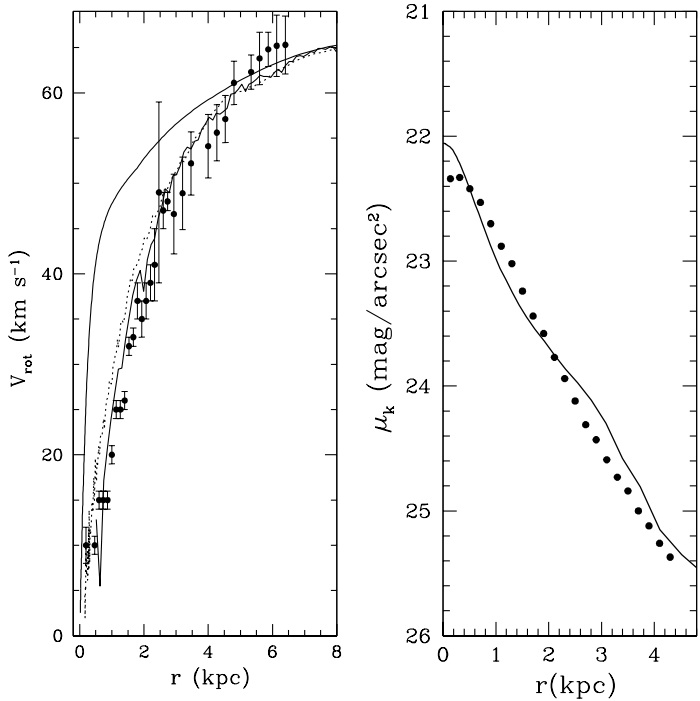}  
\caption{Comparison of the model with NGC~3109.  
{\it Left panel:} The dots with error bars show the observed rotation 
curve \citep{BlaisN3109}.  The lower full curve is the rotation 
velocity in the model as measured by artificial ``observer'' placed in 
such a way that the inclination and position angles match those of the 
real galaxy (solution of the tilted ring model).  The dotted curve 
does not include effects of projection (azimuthally averaged rotation  
velocity). The upper full curve show the circular velocity in the 
model.  {\it Right panel:} Surface brightness in the I band for the actual 
galaxy (dots) and the model (full curve).}  
\label{fig:n3109model}          
\end{figure*}

\subsection{Method}   
\label{sec:method}

The standard procedure used to fit a mass model to a galaxy requires 
one to vary the model parameters, such as the dark matter halo mass 
and concentration and the stellar M/L ratio, until the model circular
velocity reproduces the observed rotation curve.  Disk structural
parameters are fixed by the galaxy photometry. In some cases this
procedure leads to very small values for the stellar M/L not always in agreement
with the observed colors for the galaxy \citep{BlaisN3109}.  
 Alternatively, the observed colors of the stellar disk are used in combination with a
stellar population synthesis model in order to assign a stellar mass
to light ratio (M/L) \citep[e.g.,][]{McGaugh04}. There are some potential drawbacks to this
method. The procedure assumes that gas and stars in the disk move with
the local circular velocity. However, in many cases there is evidence 
of non-circular motions in the central regions of galaxies 
\citep{Simon04,Coccato04}, which are difficult to take into account.
In addition, the hot gas component can give pressure support to the
cold gas, as we find in our hydrodynamical simulations.  
Another potential complication is that parameters of the model are varied 
freely in order to reach a good fit without any guarantee of self 
consistency.  As a result, the dynamical stability of the model is questionable. In
some cases the available observational data give a maximum disk model
that is unstable to strong spiral arms or bar formation. However the
kinematics of the model is calculated as if it were axisymmetric,
stable and supported only by rotation.  This subject has received some attention by
\citet{LiaMaxdisk} and \citet{Fuchs}. It is crucial in order to
give an accurate interpretation of the observed rotation curves.  Even
if a galaxy is globally dominated by dark matter, it can develop a
strong and healthy bar \citep{Kregel2005,Lia2002,barpaper,Victor2005}.  As a result of 
the instabilities discussed above, the importance of non-circular
motions and in some cases the asymmetric drift correction can be
under-estimated assuming an exponential, infinitely thin and
axisymmetric disk at all radii. Several dwarf and LSB galaxies show
deviations from these assumptions particularly in the central regions.
The thickness of galactic disks is another issue. It is correctly
considered for the calculation of the disk potential
\citep{vandBosch01}. However, it also has a projection effect on the
kinematics \citep{Rhee} and possibly in asymmetric drift correction (See Appendix).

We adopted a different modeling procedure that accounts in a realistic
way for deviations from axisymmetry and that assures
self-consistency. The method is similar to the ones applied to high
surface brightness barred galaxies \citep{Kormendy1983,Lia1984}).
Instead of an analytic rigid model we have a high resolution N-body
realization of our system that is evolved for many disk rotation
periods.  Since gravity is a scale-free force, it is well known that
an N-body simulation can be re-scaled using two independent
variables. We use this freedom to make the models more realistic.  We use
the spatial $\alpha$ and velocity $\gamma$ scales. Thus, if $\vec{x}$
and $\vec v$ are the original coordinates and velocities, the rescaled
variables are $\tilde{\vec{x}} = \alpha\vec{x}$ and $\tilde{\vec{v}}=
\gamma\vec{v}$.  Once $\alpha$ and $\gamma$ are fixed, the mass scale
$\beta$ is defined by $\gamma^{2} = \beta/\alpha$.
 
The first step of the method is to rotate and tilt our model using the
observed galaxy position angle (PA) and inclination.  Afterward we
scale the maximum circular velocity in the model to the maximum
rotation velocity observed in the galaxy, we also change the spatial
scale in order to fit the disk radial scale length.  The next step is
to fix the bar parameters. The radial surface brightness profile is an
useful constraint on the strength and length of the bar
\citep{Elmegreen85}. We use the observed profiles of isophotes
ellipticities and PA to fix the bar orientation.  This quantity also
has a strong influence on the projected surface density.

\begin{figure*}[htb!] 
\epsscale{1.3} 
\plotone{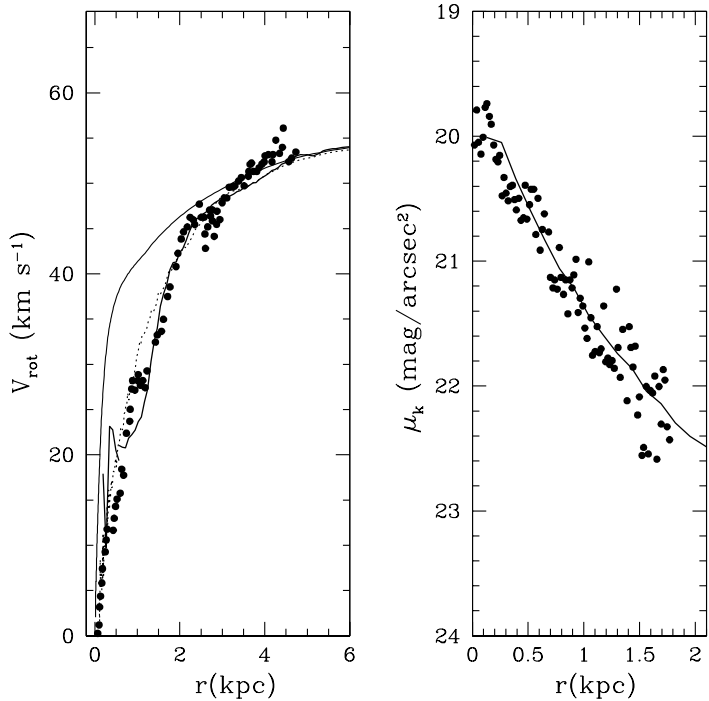}  
\caption{The same as in Figure~\ref{fig:n3109model}, but  for NGC~6822.
Observational data taken from
\citet{Weldrake2003}.  The right panel shows the unprojected surface
brightness in our model (full curve); the dots represent
K-band observations from \citet{Weldrake2003}.}
\label{fig:n6822model}  
\end{figure*}

Bar orientation is an important constraint also for kinematics. It is
well known that the effect of non-circular motions on the rotation
curve is very sensitive to this parameter \citep{Lia1984}.  This
effect is larger when the bar is orthogonal to the line of sight and
the velocities of many stellar and gas orbits are orthogonal to the
observer too. If the orientation of the bar is along the line of
sight, the galaxy would appear to have a small bulge and the
kinematics would overestimate the circular velocity. However, this
galaxy would be considered as consistent with a cuspy halo with
a very high concentration, if no modeling of the non-circular motions
is performed.

Once the disk and bar orientations are fixed, the standard tilted ring  
analysis is performed \citep{Begeman89}. We modify the spatial and velocity scales 
($\alpha, \gamma $) until we get a satisfactory match between the data 
points and our model ``observed'' rotation curve.                

We assumed an NFW density  profile  for the halo.  The virial radius is 
defined as the distance where the halo overdensity is 340 times the 
average density of the universe.  After scaling we have to re-calculate  
the value of the virial radius. This is done assuming 
the NFW profile as follows:     

\begin{eqnarray}
&\frac{3 \beta}{\alpha^{3} r^{3} 4 \pi}& \frac{M_{vir}
       F(r/R_{s})}{F(C_{\rm vir})} = 340 \Omega_{m} \rho_{cr},\\ 
       &F(x)&\equiv \ln(1+x)-x/(1+x), 
\end{eqnarray}
where $\Omega_{m}$ the average density of the matter in the Universe, 
$C$ is the halo concentration, and $R_s$ is the characteristic (core) 
radius of the NFW profile.  The same relation could be written in 
order to define a relation between the original and scaled quantities: 

\begin{equation}  
 \left(\frac{R^{'}_{vir}}{R_s}\right)^3 = 
                       C^3_{\rm vir} \frac{\beta}{\alpha^3}
                     \frac{F( R^{'}_{\rm vir}/R_s)} {F(C_{\rm vir})}.    
\end{equation}   
  
Here $R^{'}_{\rm vir}$ is the new (scaled) virial radius, while all
the other quantities have the original non-scaled values.  The halo
characteristic radius $R_{s}$ is simply multiplied by the spatial
scaling factor $\alpha$. This modifies halo concentration 
 in a non-linear
way: $C^{'}_{\rm vir} = R^{'}_{\rm vir}/(\alpha R_{s})$.

In order to mimic real observations we generated images in   
FITS format from the stellar disks in our models. The images are processed  
using IRAF. We use the package STSDAS and the task ELLIPSE in order to 
define elliptical isophotes over the image. Velocities are projected 
and integrated along the line of sight of an artificial observer.       
The isophotes are used for a tilted ring analysis that gives the 
rotation curve, the position angle  and  isophotal inclination profile 
as a result. This standard analysis assumes that the ellipticity of the 
isophotes is a measure of the corresponding ring inclination.  
Our implementation of the algorithm has as an option to assign a global inclination and  
position angle or to leave these parameters to change as a function of  
radius. The method is  described in detail by \citet{Rhee,Begeman89}.

\begin{figure}[tb!]    
\epsscale{1.1} 
\plotone{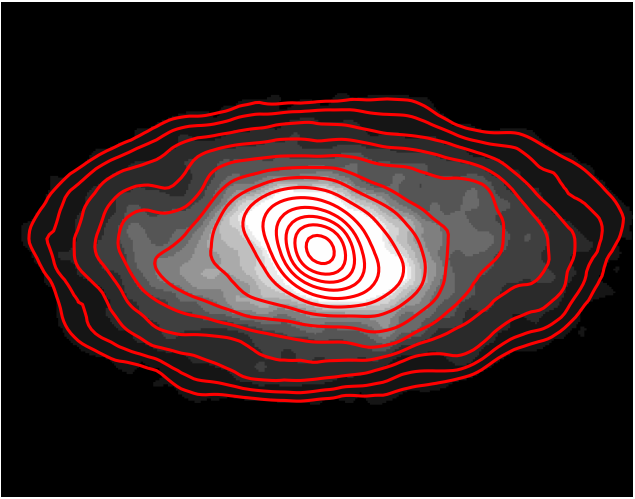}    
\caption{The surface density and the isodensity contours for our model
of NGC~6822. The projected relative position angle between the bar and
the galaxy major axis is consistent with measurements of \citet{Cioni}.
}
\label{fig:n6822iso}   
\end{figure}      

\begin{figure}[tb!] 
\plotone{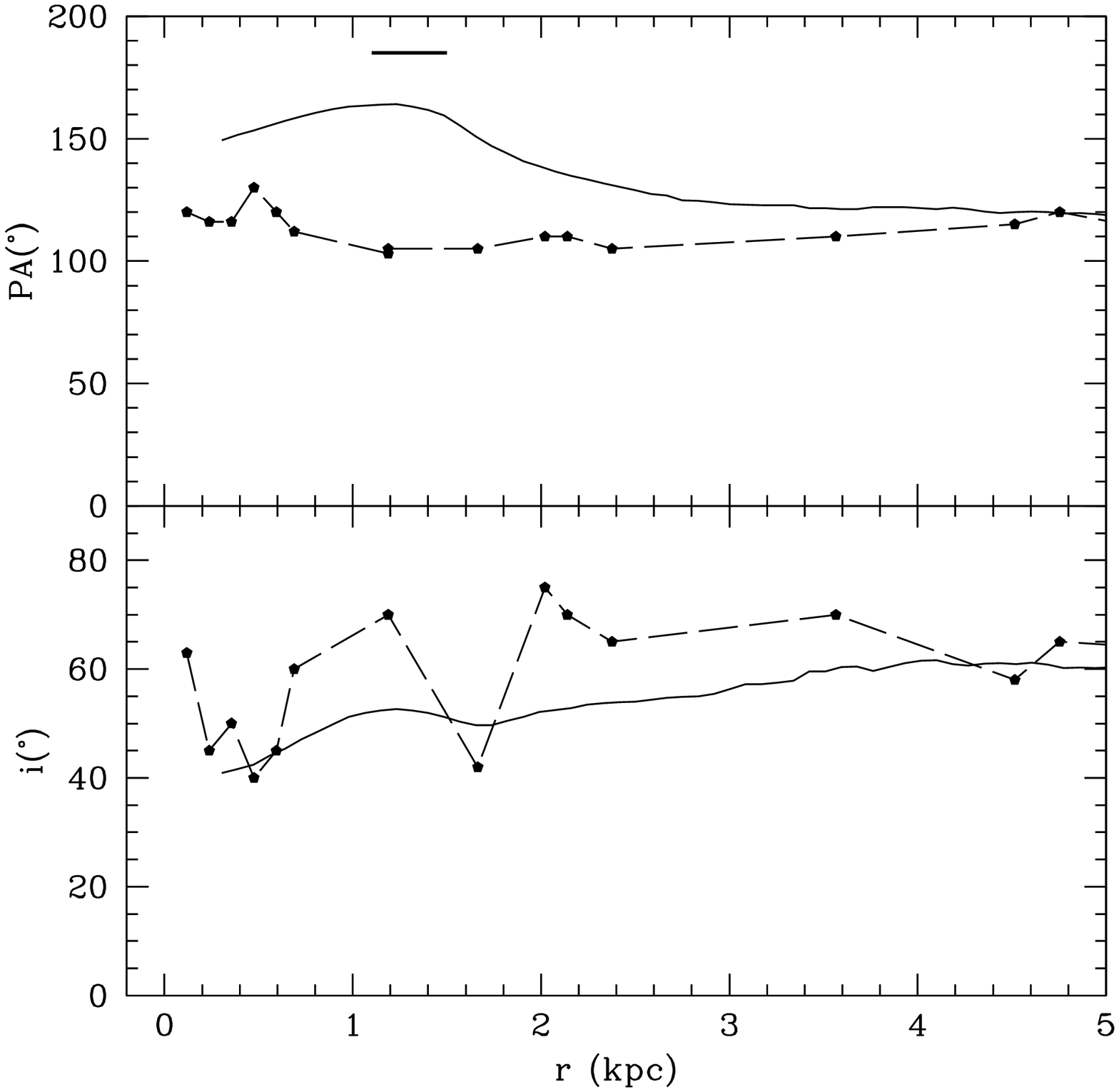}     
\caption{Position angle (top panel) and inclination (bottom panel) for
NGC~6822.  The full curves show results from our model. In the top
panel the dashed curve is from observations HI gas
\citep{Weldrake2003}.  The horizontal line at $PA=190^o$ marks the 
orientation of the optical bar \citep{Cioni,1977PHodge}.  Note that
inside 2 kpc the profile of the position angle in the simulated galaxy
shows a change with radius.  This change of $\sim 55^o$  is
consistent with the orientation of the optical bar. The bottom panel shows
the ``observed'' inclination profile calculated using the ellipticity 
of isodensity curves. The observational data are taken from  
\citet{Weldrake2003}.}  
\label{fig:N6822isophotes}  
\end{figure}

\subsection{Results} 
We analyze our galaxy model after 3~Gyrs of isolated evolution. 
At this stage the system evolves very gradually. Its bar slightly  
grows and slows down. We tried different epochs and selected one 
which matches the observations. The rotation curve fit is not  
very sensitive to a particular epoch.

\begin{figure}[tb!]
\epsscale{1.1}
 \plotone{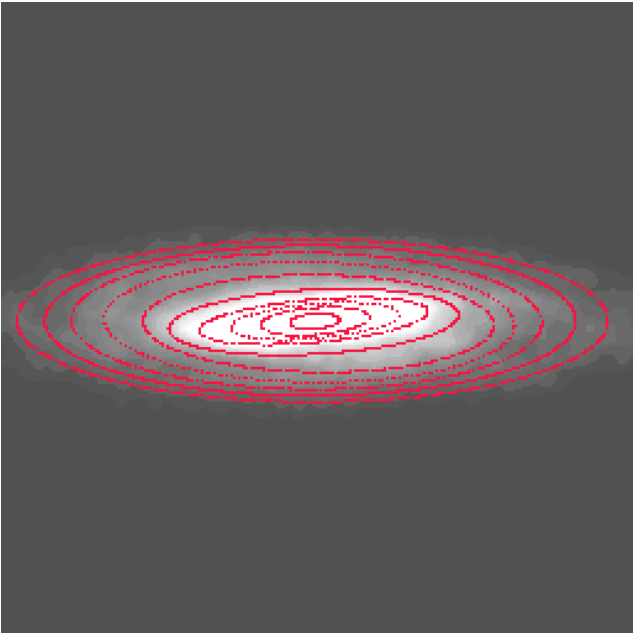}   
\caption{Surface density  for model of NGC~3109.} 
\label{fig:n3109iso}          
\end{figure}

\begin{figure}[tb!] 
\plotone{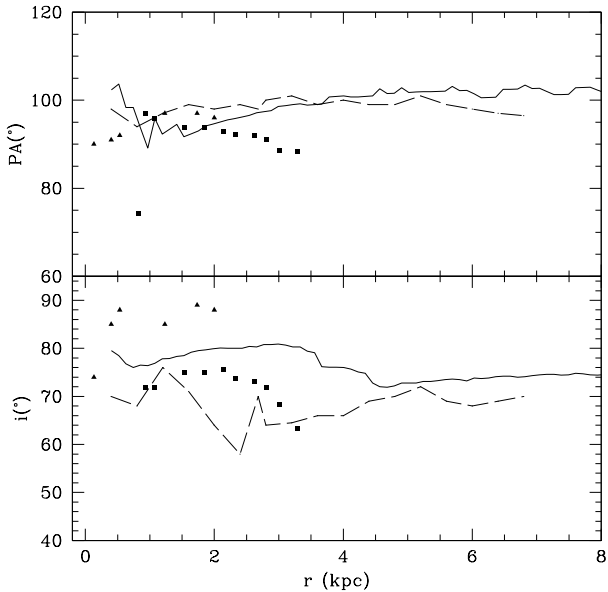}         
\caption{Isophotes in NGC~3109. The position angle and measured
inclination in the model are presented with full curve. The dashed
curve corresponds to the HI data and the squares correspond to the I
band isophotes taken from \citet{Jobin}. Triangles represent the
H$_\alpha$ observations from \citet{BlaisN3109}. }
\label{fig:n3109isophotes}
\end{figure}

Figure~\ref{fig:n3109model} shows the rotation curve of NGC~3109 and  
compares it with the azimuthally averaged rotation velocity in our  
model. We also show the surface brightness profile of the 
galaxy. Assuming that mass-to-light ratio does not change with radius,  
we compare the observed surface brightness with the surface density  
(scaled to surface brightness units) of the stellar component in the  
model. Figure~\ref{fig:n6822model} shows the same for NGC~6822.  When  
matching the models with galaxies, we adjusted the position of the  
bars in order to reproduce the gradual change in the isophote  
orientation presented by \citet{Komiyama03} for NGC~6822 and by  
\citet{Jobin} for NGC~3109.  The resulting configuration is in  
reasonably good agreement with both pictures in the central region.  
Figures~\ref{fig:n6822iso} and \ref{fig:N6822isophotes} show the  
surface density, PA and inclination angle recovered from isophotes for 
the model of NGC~6822. Figures~\ref{fig:n3109iso} and 
\ref{fig:n3109isophotes} present results for NGC~3109.  

A striking result, which is apparent in Figures~\ref{fig:n3109model}
and \ref{fig:n6822model}, is that in our models the rotation predicted
is smaller than the circular velocity. This important mismatch
between the circular velocities and the observed rotation velocities
is due to many effects.  In addition to elliptical motions and to the
asymmetric drift, there is an extra contribution coming from the fact
that our simulated observations average particles at different heights
and radii. This projection effect biases the measured velocity toward
lower values \citep{Rhee}. 
Note that the non-circular velocities in
the models are quite small and are comparable to cold gas motions in
our hydro simulation. For example, the velocity dispersion of the
stellar component in our models is smaller than 10~\kms~ for radii
larger than 1.5~kpc and it reaches 20~\kms~ at the very center for the
model of NGC~6822. The model for NGC~3109 has a central value smaller
that 30~\kms~ and the amplitude of the rms motions decreases to less
than 15~\kms~ after 1.5~kpc.

Is the mass of the disks in our models acceptable? After all, the
baryonic disk is dominant in the central 0.5~kpc in the
models. Another issue is the ratio of the disk mass to the virial
mass.  Our models have a disk-to-halo mass ratio of 2\%. This is much
smaller than the cosmological baryons-to-dark matter ratio 0.17.  Is
this a problem?  We do not think so. This is comparable to what is
observed in real galaxies. For example, for our Galaxy the ratio is
(5--6)\% \citep{KlypinZhao2002}.  In fact, the ratio of 2\% is probably realistic but on the
high side for dwarf galaxies \citep{Weldrake2003}.

The luminosity, the mass-to-light ratio, and the mass of neutral
hydrogen in the disk are other important properties of galaxies, which
a model should satisfy.  Knowing the luminosity of the galaxy and
assuming a reasonable stellar mass-to-light ratio (consistent with
observed colors), we get an estimate of the stellar mass. Observations
also provide us with the mass of neutral hydrogen. We can make a
correction for the helium  to get an estimate of the
total gas mass. Mass of molecular gas is still quite uncertain.
Observational measurements presented by \citet{Leroy2005} indicate
substantial variations in the mass ratio of molecular to neutral
hydrogen $M(H_2)/M_{HI}$. In many cases the mass of molecular gas even in
dwarf galaxies is comparable with the mass of neutral hydrogen.
The sum of the stellar and gas components gives us an estimate of the
total mass of the disk. We use this as an additional constraint for our
models.

As our fiducial model we adopt one that reproduces the rotation
 curve and the surface brightness profile and that gives acceptable
 mass of gas.  We also use observational constraints on the total disk
 mass.

\begin{table}[tb]
\caption{Parameters of $N$-body Simulation.}  
\begin{center}
\small
\begin{tabular}{lll}
\tableline\tableline\\  
Parameter& I & II  \\       
\tableline \\
Number of DM particles    &  $3.3\times 10^6$  &  $1.7\times 10^6$ \\    
Mass ratio M$_{\rm disk}$/M$_{\rm halo}$     & 0.02   & 0.01  \\   
Halo concentration C$_{\rm NFW}$   & 15. & 14.  \\         
Number of disk particles   &  $2\times10^5$  &  $2\times10^5$  \\         
Disk height-to-length ratio   & 0.085  & 0.1   \\        
 \qquad $z_{\rm d}/r_{\rm d}$  \\        
Stability parameter $Q$  & 1.2 & 1.5 \\      
Maximum Resolution   & $0.06z_{\rm d}$ &  $0.06z_{\rm d}$   \\       
 \qquad ($2\times$ smallest cell size)  & &  \\       
\tableline
\label{tab:nbody}
\end{tabular}
\end{center}
\end{table}

Observational data on luminosity and colors of the galaxies constrain
the stellar mass.  The B-R color for NGC~3109 is $\sim 0.8 \pm 0.16 $
\citep{colors}. This implies a stellar (M/L)$_B$ $\sim 0.9 \pm 0.19$
using the models presented by \citet{Bell}.  Alternatively, we can use
the maximum disk stellar (M/L)$_B \sim$ 1.5 \citep{Dutton}.  Using the
mass-to-light ratios and assuming the luminosity $L_{B}$ =$5.2\times
10^{8}\lsun$ we get the stellar mass  in the range $4.7-7.8\times
10^8\msun$. The total -- stellar plus gas -- disk mass in our model of
NGC~3109 is $2.1\times 10^9\msun$. Thus, the mass of the total gas in
our model should be $(13-16)\times 10^8\msun$.  Further assuming that
30\% of hydrogen is in molecular form and correcting for helium and
metals, we estimate that the mass of neutral hydrogen in the model is
$(6-8)\times 10^8\msun$.  These values for gas mass are somewhat
high, but, considering uncertainties, they are still compatible with
the masses of different gas components in NGC~3109 (see Table 1).  We
make the same estimates for lopsided models presented in the next
section.  Table~\ref{tab:summary} lists masses of different components
in the models.

 We now turn to NGC~6822, where the situation is more complicated. 
 \citet{Weldrake2003} gave the K-band luminosity $3.8\times 10^8\lsun$
 for the galaxy, but we consider this value as a lower limit. 
 The data covered only the central 1.6~kpc, which produces the 
 luminosity $L_K=2.4\times 10^8\lsun$ and give a very short exponential
 lenght that correspond mostly to the stellar bar.  \citet{Weldrake2003} 
 extrapolated the surface brightness profile to larger distances 
 using a one-component exponential model. For comparison with our 
 model we will use data only inside 1.6~kpc radius. In our model 
 the mass of the disk inside the radius is 
 $M_{\rm 1.6kpc}=3.4\times 10^8\msun$. Assuming 
 $(M/L)_{\rm k} = 0.5$ for the galaxy, we get stellar mass $M_{*\rm
 1.6kpc}=1.2\times 10^8\msun$. Thus, our model should have $2.2\times
 10^8\msun$ mass of gas, which is close to the estimated gas mass in
 the galaxy (see Table 1). Table~\ref{tab:summary} gives parameters of
 our model for NGC~6822. Values in parentheses are for central
 1.6~kpc. Those should be compared with the observational data in
 Table 1. Just as for NGC~3109, we assume that 30\% of the total
 hydrogen is in the molecular form, and we use factor 1.4 to convert
 from total mass to hydrogen mass.

 \subsection{Lopsided Models} 
\begin{figure*}[htb!] 
\epsscale{1.5} 
\plotone{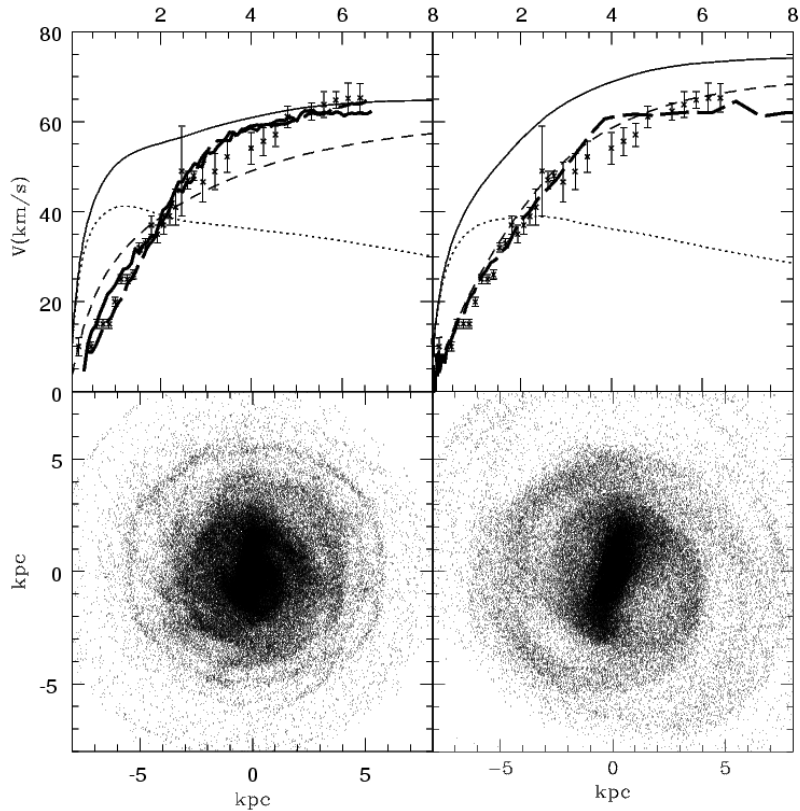}    
\caption{Effects of lopsidedness. The left (right) panels are for
mildly (strongly) lopsided model II.  In the upper panels the circular
velocity curves are shown by the top thin curves. Contributions of
baryons (dotted curves) and dark matter (short-dashed curves)  are also shown.
The thick dashed line corresponds to the rotation curve of the lopsided model. 
 As a reference in the left panel, thick solid line represents the
rotation curve for the non-lopsided version of the model.  Crosses with error
bars correspond to the observations of NGC~3109. Lower panels show the corresponding
disk particles distribution for both models. For clarity we present
particles in a slice of 300~pc thickness.}
\label{fig:newmodels}   
\end{figure*}

There are additional complications with the two galaxies besides the
stellar bar: NGC~6822 shows tidal tail-like features at large radii
and NGC~3109 is lopsided and probably warped.  These features may
result in additional non-circular motions.  In order to test these
effects, we produced lopsided models. Figure~\ref{fig:newmodels} shows
the stellar component and the rotation curves for the lopsided
models. Qualitatively results are similar to non-lopsided models. For
the lopsided models the underestimation of circular velocity is even
larger than for non-lopsided models because the centers of the
kinematic and the mass distribution have a different locations, which
leads to projection effects. In fact, the disk is not circular any
more \citep{SPhD}.  There are several spiral waves and noncircular
motions involved. The main difference with non-lopsided models is that
the lopsided models allow more concentrated DM halos. For NGC~3109 we
get $C_{\rm vir}= 11$ for mildly lopsided model and $C_{\rm vir}=
17.4$\ for the strongly lopsided model. When scaled for NGC~6822, the
models give even larger halo concentration. We conclude that
lopsidedness does not change our main conclusions. It makes them even
stronger. Altogether we have four models for each galaxy.  Parameters
of two lopsided models, which give better fits to observed galaxies,
are given in the Table~4. 

Our results are not the final models for both galaxies. We still lack
reliable observational data on the magnitude of non-circular motions
and estimates of the effects of the bar and lopsidedness on the
kinematics of these galaxies. However, we show that using existing data it is possible to
construct a family of models consistent with both cosmological
predictions and with the observed properties of the baryonic component.
Table~\ref{tab:summary} gives a summary of the properties of models
for NGC~3109 and NGC~6822.
 
\section{Discussion} 
\label{sec:discussion}    

\subsection{The Tilted Ring Analysis}      

An important consequence of our results is that the  deviations from axisymmetry 
may and often do bias the recovery of the distribution of mass in observed galaxies.
In the case of NGC~3109 and NGC~6822 these biases are enough to
explain the discrepancy between the observed rotation curves and the
theoretically expected cuspy dark matter halos.  
Following previous studies of NGC~3109 and NGC~6822, we allow the PA
and inclination to vary during the tilted ring analysis. For a
comparison, we also run the analysis fixing these quantities to the
average values measured along the disk. We find that the results are
sensitive to the assumptions made about the PA and inclination
profiles.  Our models for PA and inclination are in agreement with
observations of \citet{Weldrake2003} and \citet{Jobin}.  In our models
the changes in PA and inclination are created by the bar.

  Our interpretation of NGC~3109 and NGC~6822 as barred galaxies is 
supported by twists in the isophotes and in the iso-velocity contours 
( figures 1 and 9 of \citet{Jobin}, figures 5 and 6 of 
\citet{Weldrake2003}, figures 1 and 4 \citet{Komiyama03}).  Many other 
dwarf galaxies also have variable PA and inclination as a function 
of radius and, thus, they also may have bars.  The standard tilted 
ring analysis overlooks these variations by interpreting them as  
warps and assuming that the material out of the disk plane still moves
with the local circular velocity which is not obvious \citep{wong}. 
While warps are definitely frequent in galaxies, warps are  
believed to be produced by tidal interactions with a companion, a   
satellite or a triaxial dark matter halo.    
Those interactions should {\it increase} with the distance    
from the center of the galaxy. Yet, the changes in PA and inclination    
in NGC~3109 and NGC~6822 happen in the central $\sim 1$~kpc regions of     
the galaxies.  In the case of NGC6822 the presence of the bar as a feature     
in the stellar component is evident from the maps of the youg stellar component 
presented by \citep{dBW2006} wich is consistent with the central change of PA in the
neutral hydrogen and out of phase with the tidal tails.  In the case of NGC 3109   
the stellar disk seems unperturbed  out of one kiloparsec,  
suggesting that the change in position angle is created by a bar or an oval 
distorsion \citep{Jobin}.  
However both possibilites: warps or oval distorions do not exclude each other, making the 
situation complicated.  The ultimate answer will require a detailed modeling of  
the non-circular motions and the projection effects with methods 
similar to the ones discussed by \citet{Schoenmakers}, and it will     
also require the corrections to the kinematics discussed in the APPENDIX.

\subsection{Congruency with Cosmology}

\begin{planotable}{lccccrr}      
\tablecolumns{7} 
\tablewidth{0pt}  
\tablecaption{Models of NGC~3109 and NGC~6822}  
\tablehead{\colhead{Parameter} & \colhead{NGC~3109}  & \colhead{NGC~3109}  & \colhead{NGC~6822}  & \colhead{NGC~6822} \\   
  &   & \colhead{LOPSIDED}  &   & \colhead{LOPSIDED} \\   
 }    
\startdata  
Virial mass \hskip 1em $M_{\rm vir}, \msun$ & $8.1\times 10^{10}$ & $4.7\times 10^{10}$ & $3.4\times 10^{10}$ & $1.93\times 10^{10}$  \\    
Halo concentration \hskip 1em $C_{\rm vir}$   & 14 &  17.4     & 22        & 41\\    
Disk mass \hskip 1em $M_{\rm disk}, 10^8\msun$& $21.0$ & 15.5  & 11.0      & 6.39  \\      
Stellar mass M$_{\rm stars}, 10^8\msun$  &$4.6-7.8$&$4.6-7.8$  & 1.9 (1.2) & 1.9 (1.2) \\ 
Neutral hydrogen mass, $10^8\msun$       & $6-8$   & $6.3-6.9$ & 4.3 (1.0) & 2.1 (1.1) \\            
Total gas mass, \hskip 1em $10^8\msun$   & $13-16$ &$7.7-10.9$ & 9.1 (2.2) & 4.5 (2.4)  \\           
Stellar (M/L)$_{B}$   &  $0.9-1.5$ &     $0.9-1.5$ & \\   
Stellar (M/L)$_{K}$   &      & & 0.5 & 0.5  \\  
Bar Orientation, degrees   &30  &30 &  45  & 45  \\ 
\label{tab:summary}    
\enddata  
\tablecomments{{\it NGC~6822:} Values in parentheses are for inner 1.6~kpc region.}
\end{planotable}

\begin{figure*}[htb!]      
\epsscale{1.5}  
\plotone{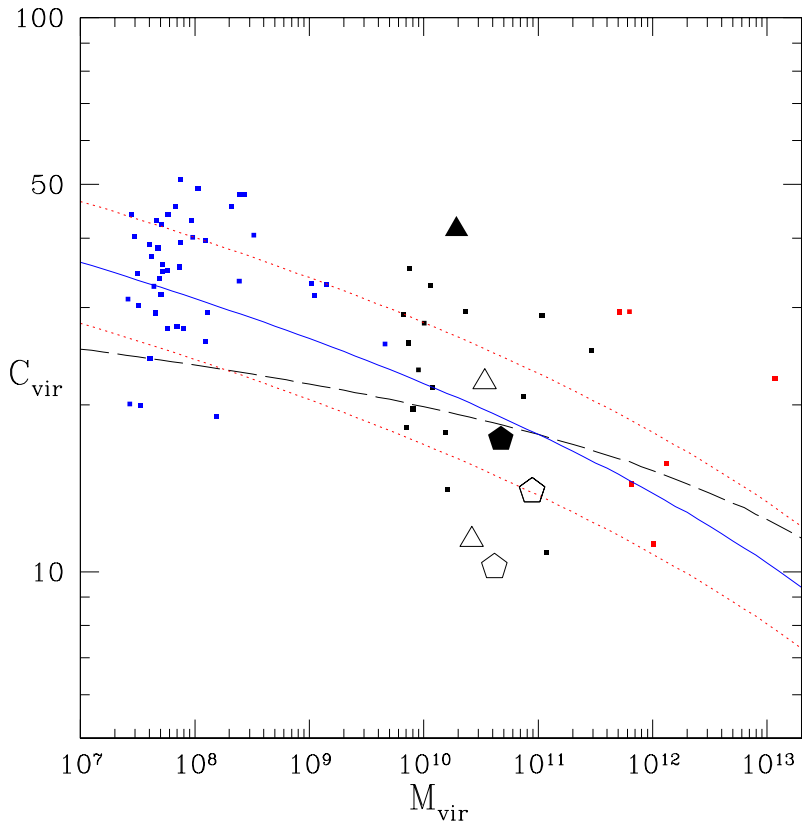}            
\caption{Comparison of halo concentrations for three of our models with 
cosmological predictions. Virial concentrations of our models are shown by large symbols.  Models represented by open 
symbols are not lopsided. Pentagons corresponds to NGC~3109 models, 
triangles are for NGC~6822.  
The small squares show the virial  
mass and concentration measured for halos in cosmological N-body  
simulation (Col\'in et al 2004). The central thick line is the model 
for the relationship between virial mass and concentration proposed by 
\citet{Bullock01},  thick line is the mean value and the dotted 
lines indicate the 1~sigma scatter. The dashed line is the model from 
\citet{Eke2001}. The halos in all our models are comparable to those found in cosmological 
simulations.} 
\label{fig:cvir-mvr}      
\end{figure*}

Our models reproduce the observed rotation curve as well as some of 
the disk structural properties of both galaxies.  We now evaluate  
whether the dark matter halos in our models are consistent with  
cosmological predictions.  Figure~\ref{fig:cvir-mvr} shows the virial 
concentration and mass for a population of dark matter halos formed in 
cosmological collisionless N-body simulations. The amplitude of 
fluctuations for these simulations was normalized to have 
$\sigma_8=0.9$. Recent results of the third year WMAP revision
favor smaller amplitude 
$\sigma_8\approx 0.75-0.80$ \citep{WMAP3}, which decreases the 
predicted average central density of halos. If LSB galaxies are hosted by
relatively low concentration halos \citep{Bailin05}, 
cosmological predictions for concentration will be 
reduced almost by a factor of two \citep[Maccio et al. in  
preparation]{Alam02}. This rescaling will favor our non-lopsided 
models. However, given that the effect of adiabatic 
contraction or any possible evolution of the dark matter halo after  
the disk formation is not included in the relationship between 
halo mass and concentration, we still compare with the high $\sigma_8$ 
prediction.

The rotation curve and the disk structure of both galaxies -- NGC~3109 
and NGC~6822 -- are consistent with a range of cuspy dark matter
profiles.  The four models (two non-lopsided and two lopsided) for
NGC~3109 are consistent within the uncertainties with the
observational estimates of the disk mass. Two of them are within the
one-sigma scatter of the average concentration predicted for halos of
the corresponding mass. In the case of NGC~6822 the strongly lopsided model II
has a concentration well above the average prediction. The  
non-lopsided version of model II has a relatively small value of the
concentration and it is consistent with models discussed by \citet{Carigi06}. 
In this respect it is remarkable how strong is the effect of 
lopsidedness, and it also suggests that systematics related with lopsidedness 
and bars can increase the scatter  of concentrations measured using rotation curves.     

Yet, not always it produces visible differences. Figure~\ref{fig:newmodels} shows the
distribution of disk particles in lopsided models. Although  the disk shows the
asymmetry, the rotation curves are very regular.  We can not decide 
which model -- lopsided or not -- is better without more accurate  
observational data on non-circular motions in NGC~6822. In any case,
in all our models the dark matter halos are compatible with
predictions of cosmological models. 

Figures~\ref{fig:newmodels} and \ref{fig:hydrorot} present the
circular velocity decomposition, showing the individual disk and halo
contributions, as well as the combined amplitude. The disk dominates
in the central region while the dark matter is dominant in the outer
regions.  This is typical for barred galaxies
\citep[e.g.,][]{Lia1984}. This is an important difference with more 
naive traditional models where the disk is treated as  thin and 
axisymmetric and where the dark matter dominates at all radii.  Bar 
lengths in our models are around 2.4~kpc and 2.0~kpc. The length of bars is
reflected in the stellar kinematics, as features in the rotation
curves. Still, those features may be difficult to detect because their
manifistations are very sensitive to the assumed inclination and
position angle that are used in the tilted ring analysis.  Because
bar lengths in our models are comparable to the outer disk radial scale
length, there is a correlation between the size of the region affected
by the bars and the disk scale length.

\subsection{Generality of our Results}   
The systematic effects discussed in this paper (see also \citet{Rhee})
can affect high accuracy studies of the gravitational potential in 
galaxies regardless of the specific structure of the dark matter halo 
or the detailed form of the gravity law. These systematic effects 
are more exacerbated for non-axisymetric (barred or lopsided) disks  
however, they are  also present in axisymetric disks.

Our results regarding NGC 3109 and NGC 6822, add to the recent 
arguments of \citet{Rhee},\citet{Simon04} and \citet{Spekkens05} that  
some well studied galaxies and the galaxy population on average are 
consistent with cuspy dark matter halos.  Galaxies in our study have 
been considered impossible to reconcile with $\Lambda$CDM cosmology  
and previous analysis motivated to explore the modification of dark  
matter properties or gravity at low acceleration values, or 
alternatively the galaxies were considered an indication of a crisis  
for CDM cosmology. \citep{Moore2001,SalucciDDO47,AR01,Colin02}.

It is known that both galaxies cannot be treated as simple 
 axisymmetric disks \citep{deVaucouleurs91,CarignanN3109,1991PHodge}. 
 We show that these asymmetries are enough to bias previous analysis 
 toward models with central flat cores.  

Our hydrodynamical simulations indicate that because of their low
mass, the kinematics of the cold gas in dwarf galaxies is likely
affected by pressure gradients triggered by stellar feedback.  Once we
include these effects in the analysis, the galaxies can be consistent
with cuspy dark matter halos.  Many of the most discrepant cases like
NGC~3109, NGC 6822, IC2574, LMC, and probably DDO47 show bar-like
structures, lopsidedness and a large number of shells in their
ISM. Our conclusions should be applicable at least to some of them.
Finally, we found that estimations of the gas velocity dispersion at
scales of tens or a few hundreds of parsecs may underestimate the
amplitude of the asymmetric drift correction for galaxies that show
non-circular motions.

It has been recently argued that including the adiabatic contraction
and a realistic stellar M/L motivated by stellar population models 
makes the baryonic mass contribution in the center of LSB galaxies to 
be important\citep{McGaugh04} . This seems to be at odds with common expectations
that LSB galaxies are totally dominated by the dark 
matter. We emphasize that often those expectations are not based on  
any solid arguments.  Our models support the idea that in the central 
regions of LSB galaxies the baryons cannot be neglected. The important
implication of this picture is that the galaxies are dynamically
unstable to bar and spiral arm formation, creating non-circular
motions that lead to the underestimation of the circular velocity.  

As a possible  solution to the cusp problem of $\Lambda$CDM    
cosmology, it has been suggested that the angular momentum transfer 
into dark matter halos by galactic bars and satellites can trigger the  
formation of a flat core  \citep[e.g.,][]{HB2003,Ez04}.   
Currently there is a discussion about the efficiency of such processes.   
We can not discard these mechanisms, and kinematic studies of galaxies
might help to set constraints to the potential evolution of the 
dark matter profile. However, given that the effects created by bars --  
lopsidedness, projection effects and pressure gradients -- were not 
considered before, the evidence for a flat core may be considerably weaker 
than it has been assumed.

A crucial test for our models of NGC~3109 and NGC~6822 would be a
measurement of stellar random motions and non-circular motions in the
central region of these galaxies.  \citet{BlaisN3109} present the
H$_\alpha$ velocity field for NGC~3109 in their Figure~2. It is
possible to see fluctuations in the velocity field of about 20~\kms.
The kinematic fluctuations detected in H$_\alpha$ are
consistent with our models, given that the kinematics of ionized gas
is likely correlated with the stellar kinematics
\citep{Hunter2002,Rhee}. However, a direct measurement of the stellar
kinematic would be a better test. \citet{N6822polar} present 
stellar kinematics in the spheroid of NGC 6822 that can potentially
be combined with the disk kinematics to set tigther constraints 
on its dark matter halo.

\section{Conclusions}           
\label{sec:conclusions}         

In this paper we present self-consistent numerical models for NGC~3109 
and NGC~6822 that reproduce their observed rotation curves and some of 
the structural properties of their disks.  Our high resolution models  
include a number of effects present in real dwarf galaxies like non 
axisymmetric dynamics and star formation induced gas turbulence.

Our main conclusions are:  
 
 A detailed modeling of the baryonic component and a detailed analysis 
 of the kinematics of the tracer population shows that a constant 
 density core for the dark matter halo is not required to successfully 
 reproduce the rotation curves of even the most problematic cases for 
 $\Lambda$CDM, like the galaxies NGC 3109 and NGC 6822. On the contrary, the  
 observed rotation curves of these galaxies are consistent with 
 predictions from $\Lambda$CDM and they do not provide undisputed evidence for 
 a central flat core. 

Our models are centrally dominated by baryons in agreement with observational 
estimates of stellar M/L ratios based on stellar population models and the colors of 
dwarf galaxies  \citep[e.g.,][]{Simon04}. In many situations we find that  
models centrally dominated by baryons are unstable to bar formation, 
which implies the presence of several biases in the interpretation 
of the rotation curve. It is therefore important to test the 
self-consistency of the models used to reproduce galactic rotation 
curves.

Estimations of  non-circular motions and  rotation velocity  
should carefully take in account projection effects, particularly
for not axisymetric disks, otherwise corrections similar to the 
asymetric drift will not recover the circular  velocity.

It is possible that other effects such as the halo triaxiality  
also bias estimates of the amplitude and shape of  
rotation curves in non-barred galaxies \citep{hayashi04}. If the galaxy is  
barred, the halo is still prolate in the center. In this case the  
central halo shape is coupled to the baryonic component and the triaxiality is  
mutually driven by the disk and the halo component \citep{Colin06}. 
If the galaxy is lopsided, the underestimation can be more 
severe. If the circular velocity  is  small ($\leq 20-50$ km/s), as it happens  
in the smallest dwarf irregular galaxies and in the central regions of large 
disk galaxies, the pressure gradients triggered by star  
formation and feedback also contribute to the underestimation of the   
circular velocity even if the galaxy is axisymmetric.

Accurate measurements and correct treatment of variations in the   
position angle and inclination as a function of radius along the   
galactic disk are in some cases critical to set constraints on the 
mass distribution using rotation curves. If present (as is often the case), 
those variations may indicate a bar or an ellipsoidal  
distortion.  Different answers can be obtained adopting different    
interpretations for the PA and inclination angle profile,  if the  
galactic disk is not axisymmetric \citep{projections}.  Here we show that a method 
of analysis that includes the possibility of non-axisymmetric disks is 
necessary.

In summary, the main lesson of our modeling of mass distribution in     
dwarf galaxies is that those galaxies, which were considered as simple dark matter 
dominated objects with thin cold rotating disks, are actually quite 
complicated. Efforts to treat them in an overly simplified way result in 
substantial errors. We find that those errors always {\it underestimate}  
the true density in the central parts of galaxies.  If confirmed, 
our results suggest that there is no contradiction between 
the observed rotation curves in dwarf galaxies and the cuspy central 
dark matter density profiles predicted by the Cold Dark Matter model.       
Accurate constraints on halo density profiles will require the modeling 
of the systemtic effects discussed in this paper.

\acknowledgments   
O. Valenzuela acknowledges  A. Bosma, E. Athanassoula,J. Gallagher,
J. Navarro, M. Tavares,  J. Simon, V. Avila-Reese, S. Faber, G. Lake, 
F. van den Bosch, L. Carigi, P.  Col\'in, G. Gentile, A. Maccio and
many others for valuable conversations.   
F.Governato is Brooks fellow. O. V. acknowledges support by the  
NSF ITR grant NSF-0205413. A.K. acknowledges support by the  
NSF AST-0407072 grant to NMSU.  Computer simulations presented in this 
paper were  performed  at the National Energy Research Scientific 
Computing  Center at  the  Lawrence Berkeley National Laboratory  
and also at the Pittsburgh Super Computer Center, using lemieux.

\onecolumn
\section{APPENDIX: Motion of cold gas and the Asymmetric Drift Correction.}

\label{sec:ADC} 

In this section we give more detailed analysis of hydrodynamical 
simulations. We also try to answer some questions regarding gas motion 
in those models. It is important to clarify that in most cases we will 
consider the rotation curve measured in the disk plane or the true rotation  
as in \citet{Rhee}. An observer should consider the projection 
effects in order to recover the circular velocity.
These projection effects are particularly considerable if the disk is not axisymetric.  

We start with an explanation of why the rotational
velocity of the cold gas is significantly lower than the circular
velocity as demonstrated by Figure~\ref{fig:hydrodens}. In order to
disentangle different effects, we make two runs of model H2 wich is
axisymetric and dark matter dominated at each radii: one model includes
only hydrodynamics and cooling and the other includes in addition
star formation and feedback.  Model H2 does not form a bar after
one gigayear. So, by comparing these runs we can study the effect 
of stellar and supernovae feedback in the absence of a bar or an 
oval distorsion.     

\begin{figure}[tb!]   
\epsscale{1.0}     
\plotone{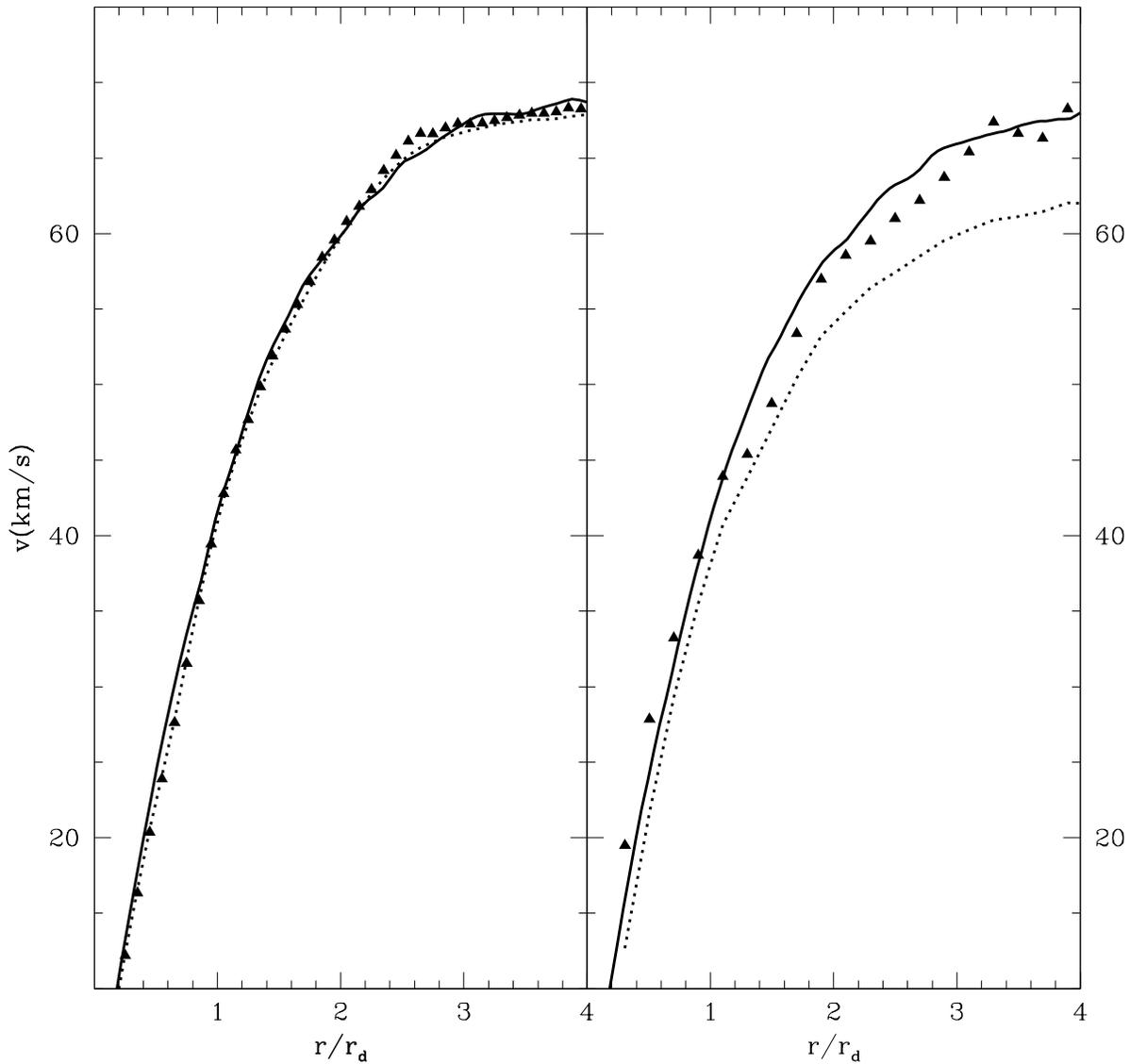}          
\caption{ The rotational velocity of cold gas (dotted curves) and
circular velocity (full curves) in hydrodynamical simulations H2, which do not have a bar.
Triangles show results of recovery of the circular velocity from the rotation curve using the
asymmetric drift correction.  The left panel shows the model with gas
cooling but without the star formation. The cold gas rotates very fast
and closely follows the circular velocity.  The right panel is for the
model with the star formation and with the feedback.  
Gas rotates systematically slower as compared with the circular 
velocity.} 
\label{fig:NoBar}  
\end{figure}

The left panel in Figure~\ref{fig:NoBar} shows different velocity 
curves for the model H2 without the feedback.  It is clear that in 
this case the cold gas has rotation velocity very close to the 
circular velocity. This test tells us that numerical effects are not 
responsible for the disagreement between the rotation and the circular  
velocities in the model H1. The right panel in Figure~\ref{fig:NoBar} 
indicates that the feedback is partially responsible for the 
discrepancy: in the model with the feedback the cold gas rotates 
systematically slower than in the no-feedback model.   The  
difference between the rotation and the circular velocities is about   
10\% at most radii but it is different inside one radial exponential  
length. The absolute value of discrepancy is smaller than in the case of the    
barred model. Still, the effect is definitely present, and it will 
affect more the slope of the recovered density profile than the average 
value of the central density. The ten percent   
difference seems to be small, and one would be tempted to neglect
it. Yet, it should not be neglected because another effect goes in the
same direction and the cumulative effect of two seemingly small
contributions is substantial. \citet{Rhee} argued that due to the
inclination, the ``observed'' rotation velocity is slightly smaller
than the true velocity. The effect is relatively simple: for a disk
with a finite thickness the average 3D radius at which the rotation is
estimated is slightly larger than the projected distance. A
combination of a rising rotation curve and a projection of the
rotation velocity to the line-of-sight results in a smaller ``observed''
rotation velocity. For a typical inclination of $\sim 70\degree$\ the
effect is about 10\% (see figs 5 and 7 in \citet{Rhee}). The
combination of those two effects -- projection and feedback -- reduce
the rotation velocity by at least 20\%.  When we use eq.(\ref{eq:rho}) to
estimate the total density, we find that the ``observed'' density is
only 0.64 of the true density.  It is quite remarkable that 10\%
effects, which normally would be ignored, produce a factor of two
underestimation of the total density.  This is also a lesson: when
dealing with rotation curves every effect on the level of $\sim 10\%$
must be included. 

In order to understand other reasons for the disagreement between the
large circular velocity $V_c\equiv\sqrt{G M(<r)/r}$ and the small gas
rotational velocity $V_{\rm rot}$, we consider and estimate different
effects related to random velocities and pressure gradients.  
 We can write an equation relating the circular and the
rotation velocities of the gas, if we assume that gas with pressure $P$
and temperature $T$ has random bulk velocities with radial dispersion
$\sigma_r^2$ and velocity anisotropy $\beta\equiv
1-\sigma_\phi^2/\sigma_r^2$. In this case the analog of the asymmetric drift correction
 is given by the following expression:
\begin{equation}    
 V_c^2 = V_{\rm rot}^2 - \sigma_r^2
            \left( \frac{\partial\ln\Sigma}{\partial\ln R} 
             +  \frac{\partial\ln\sigma^2}{\partial\ln R} - \beta \right) +
           \frac{kT}{\mu m_H}\left( \frac{\partial\ln\Sigma}{\partial\ln R} + 
           \frac{\partial\ln T}{\partial\ln R}   \right)
\label{eq:adcora}.
\end{equation}       
\noindent Here $\Sigma(r)$ is the surface density of the gas, $\mu$ is
the molecular weight, and $m_H$ is the hydrogen mass. This equation is
the Jeans equation in cylindrical coordinates with additional pressure
terms. This equation is already a simplification because, as it is
usually done, we neglect effects of a possible tilt of the velocity
ellipsoid: $\partial\langle v_rv_z\rangle/\partial z=0$.  We also
assume that the gas surface density is proportional to the gas density
in the plane of the disk.  Note that in spite of the fact that gas has
isotropic pressure, gas bulk motions in the presence of a bar or an
ellipsoidal distorsion typically are not isotropic. In 
all our models the rms velocities in the disk plane are about equal:
$\sigma_r\approx \sigma_\phi$. This implies that $\beta\approx 0$. In
the model with a bar the rms velocity perpendicular to the disk is
about twice smaller than in the plane of the disk: $\sigma_z\approx 
0.5\sigma_r$. As a result of this anisotropy in the kinematics 
the observational estimation of the velocity dispersion in the disk 
plane will be affected by projection effects.

\begin{figure}[tb!]   
\epsscale{1.11} \plottwo{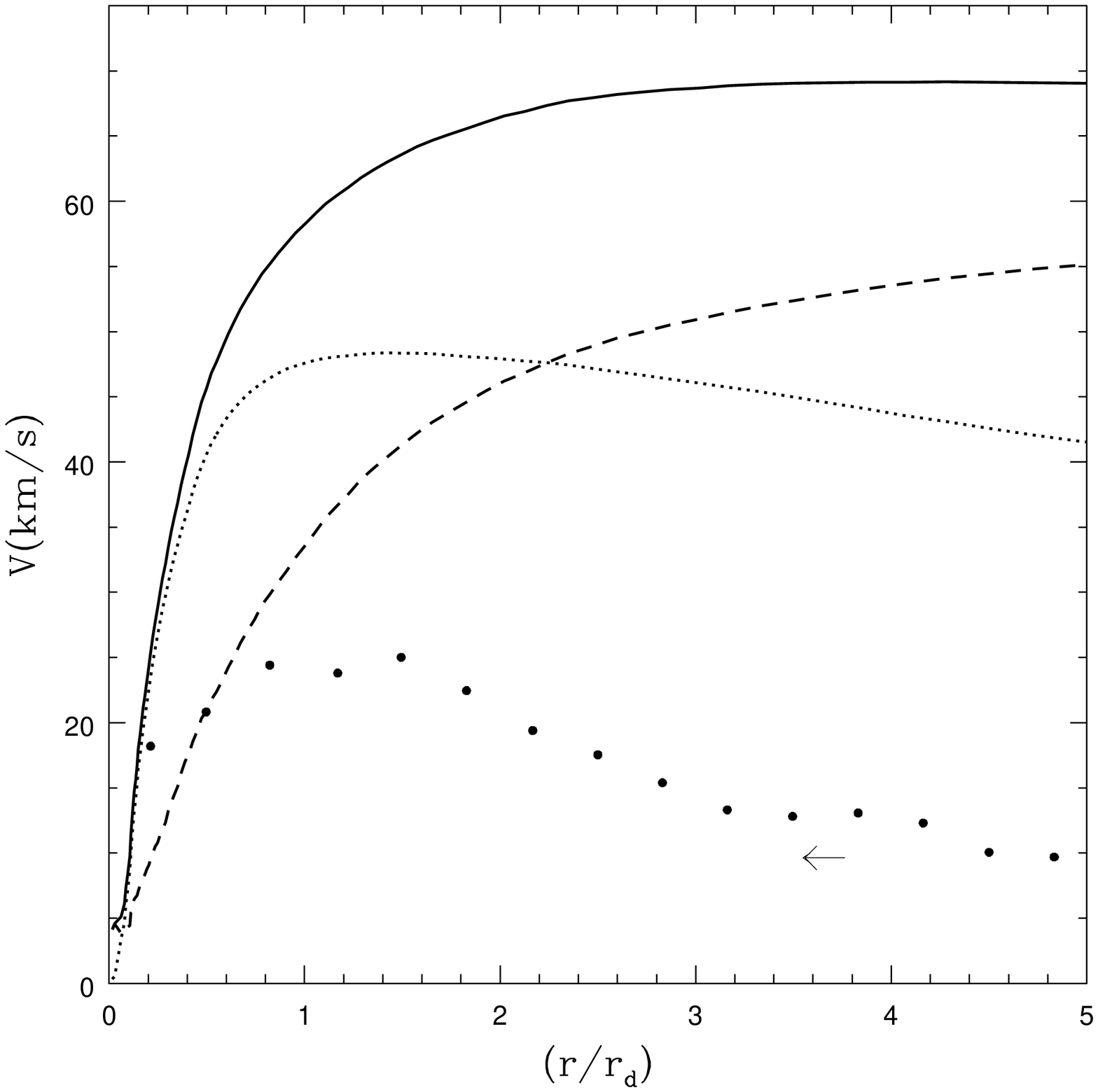}{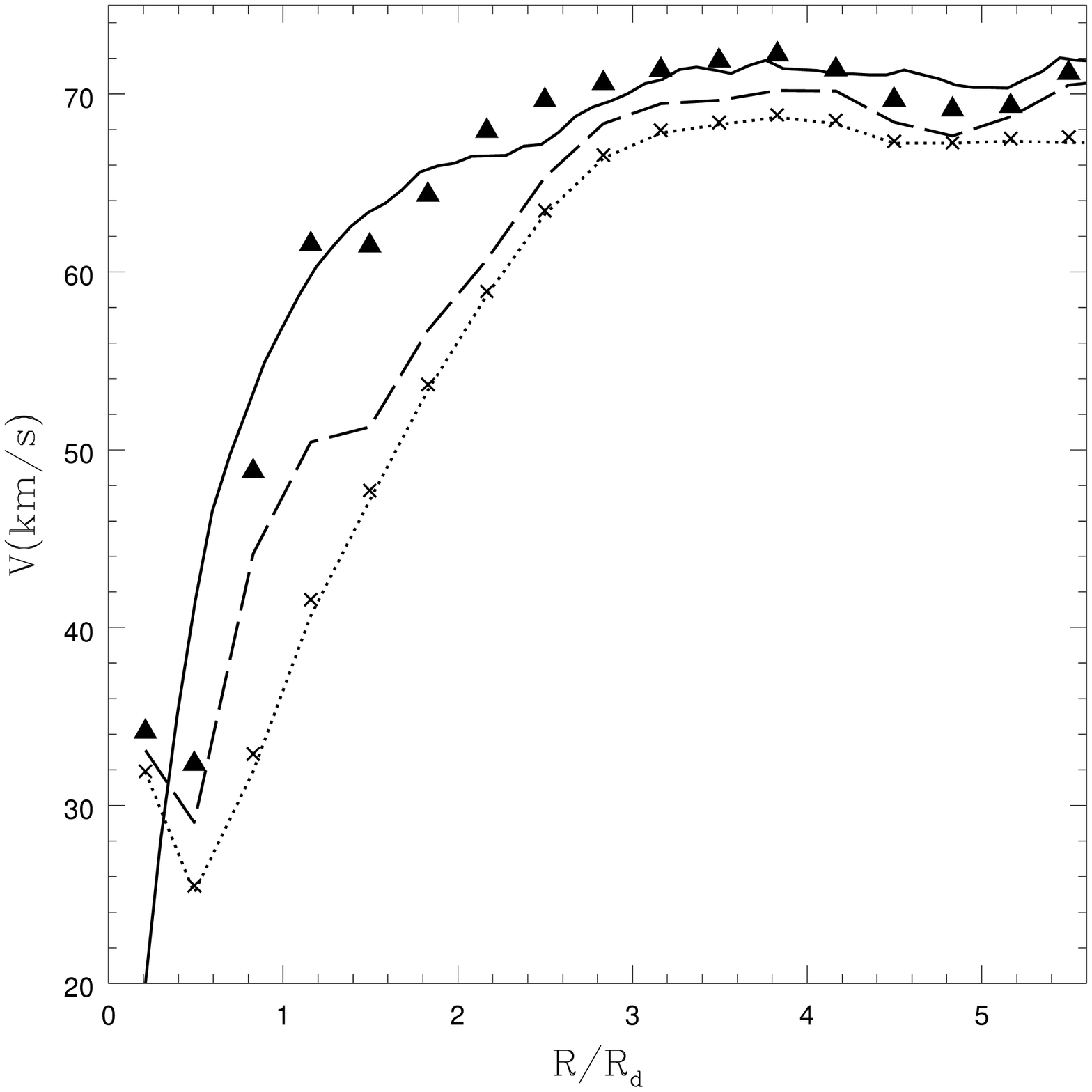}
\caption{{\it Left:} Velocity curves in the hydrodynamical simulation
 H1, which has the star formation and a bar. The full curve presents
 the circular velocity $\sqrt{GM/r}$.  The contribution of the dark
 matter halo is shown by the smooth dashed curve. The dotted curve
 presents the contribution of the stellar disk. In the central region
 the disk dominates, while the outer part is dominated by the dark
 matter.  The circles show the rms radial velocity of the cold gas
 calculated in concentric rings. The arrow indicates the line-of-sight
 velocity dispersion measured by an observer, which has the resolution
 of 100~pc and sees the ``galaxy'' inclined by $40\degree$.
Note the difference between the true and observed rms velocities.  {\it
Right:} Recovering circular velocity curve in the same model. The full
curve is the true circular velocity in the plane of the disk. The
rotation velocity is shown by the dotted curve. Crosses show circular 
velocity recovery results
of asymmetric drift correction assuming only a constant rms velocity of the
gas  $6~\kms$. The dashed curve is the recovery results when the 
only correction term is the gas pressure gradient $(kT/\mu
m_H) (d\log\Sigma/d\log r)$. The triangles present results of
circular velocity recovery with all the terms included. }
\label{fig:hydrorot}  
\end{figure}

Left panel in Figure~\ref{fig:hydrorot} shows contribution of the
stellar and the dark matter components to the circular velocity. Here we use spherical
approximation to estimate the circular velocity $V_c\equiv
\sqrt{GM/r}$. It
also shows the rms radial velocity $\sigma_r$ of the cold gas, this estimation
of the velocity dispersion includes small scale random velocities as well
as large scale bulk motions created by the bar. We used this estimation of  $\sigma_r$
in  eq.~(\ref{eq:adcora}). Yet, this rms velocity is not what
a typical observation would produce. Depending on how the rms velocity is
measured in real observations the results can be significantly
different. The observed point-by-point rms velocity $\sigma_{\rm los}$
is a mixture of two contributions: $\sigma_\phi$ and $\sigma_z$. For
example, for an inclination of 45\degree\ the ``observed'' $\sigma_{\rm
los}\approx 0.8\sigma_r$, where we assume that
$\sigma_r\approx\sigma_\phi\approx 2\sigma_z$. Even $\sigma_{\rm los}$  is
difficult to measure because of the clumpiness of the ISM in dwarf 
galaxies and because of the noise in measuring of the velocities. As
the result, observers typically report and use average rms velocity
for the whole disk. This additionally reduces the value of the rms
velocity because a large number of the pixels are in a region
out of the bar or oval distorsion. If we apply these 
procedures to our model, we get $\sigma_{\rm los}\approx (10-12)~\kms$. 

 Results of recovering the circular velocity from the rotation curve
 are presented in the right panel of Figure~\ref{fig:hydrorot}.  To
 find the circular velocity we use the gravitational acceleration $g$
 in the plane of the disk $V_c=\sqrt{rg}$.  Just as many observers
 typically find, we also conclude that the standard asymmetric drift
 correction makes very little effect. When applying the standard
 correction, we use the true log-log slope of the surface density of
 the cold gas and assume a constant rms velocity. In the Figure
 ~\ref{fig:hydrorot} the crosses are for the standard correction with
 $\sigma_r =6~\kms$. Changing this value to 10~\kms\ does not make any
 difference. For example, at $r=r_d$ the rotation velocity is $v_{\rm
 rot}=37~\kms$, the circular velocity is $v_{\rm circ}=58~\kms$, and
 the log-log slope of the surface density is -1.8.  If we use $\sigma
 =10~\kms$, the correction gives $39~\kms$: only a 2~\kms\ difference!
 When we use the {\it true} local rms velocity $\sigma_r=22~\kms$, we
 get significantly larger corrected velocity of 47.5~\kms. Still, it
 does not bring us to the circular velocity, but it is getting much
 closer.  At these small distances there is another factor, which
 should be included in the correction: the pressure gradient. Effect
 of only the pressure gradient is shown as the dashed curve in the
 plot. It makes an impact at all radii, but at $r> 1.5r_d$ it is a
 small contribution. Yet, at smaller distances it is important. At
 those distances the mass-weighted temperature of the gas in the plane
 of the disk is $\approx (20-30)\times 10^4$~K. Making all the
 corrections -- for true $\sigma_r$ and for pressure gradient -- gives
 the corrected value of rotation velocity to 57~\kms, which 
 practically is the true circular velocity. Triangles in the plot show
 results of application of the corrections at different radii. Indeed,
 we recover the circular velocity.

\begin{figure}[tb!]  \epsscale{0.9} \plotone{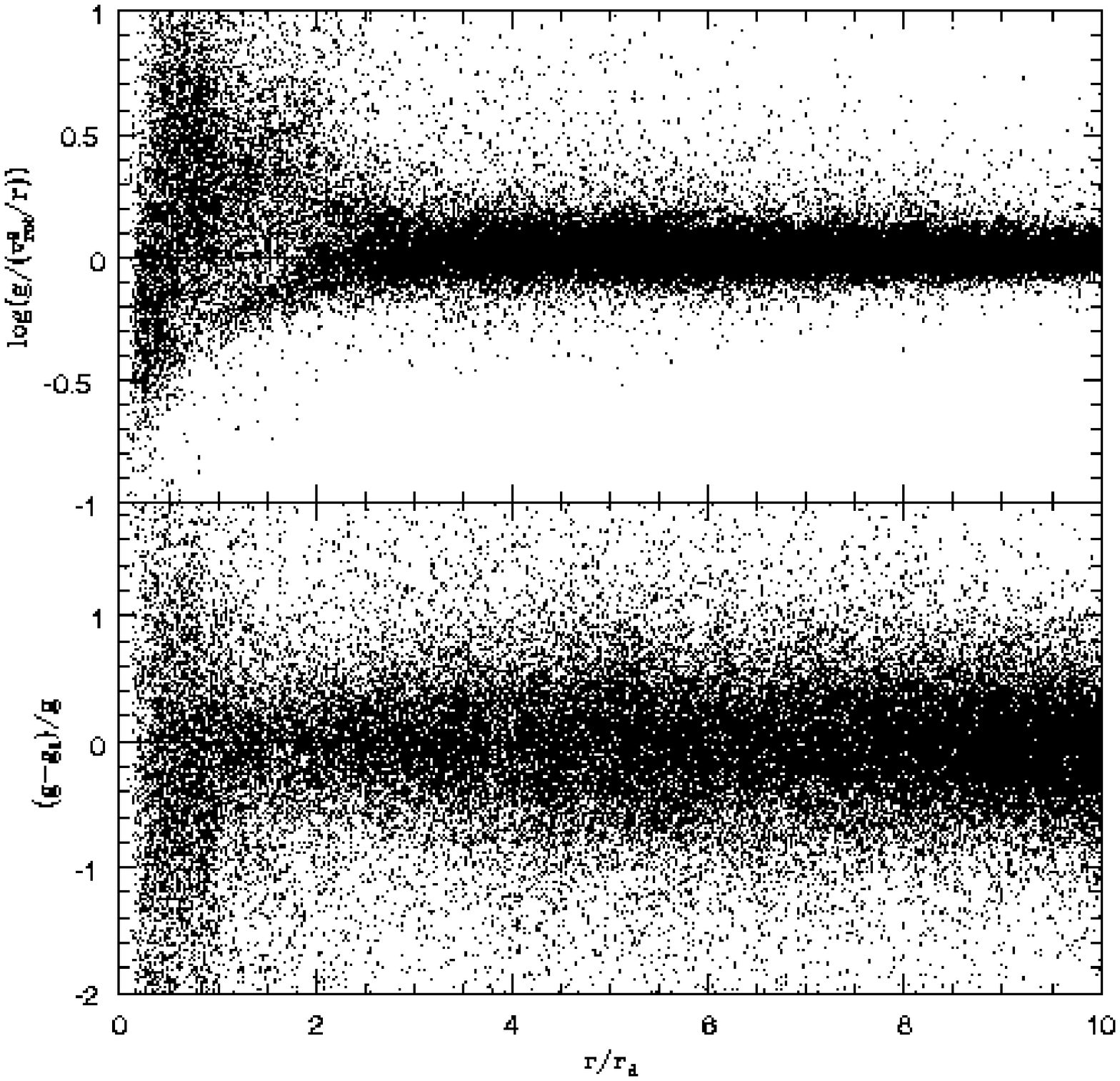}
\caption{Accelerations of individual cold gas particles in the barred model. The top panel
shows the distribution of the ratio of radial gravitational
acceleration $g$ calculated by the GASOLINE code to the centrifugal
force $v^2_{\rm rot}/r$, where $v_{\rm rot}$ is tangential velocity component
of each gas particle. At large distances the ratio is close to unity
indicating that fluid elements are close to being rotationally
supported. At $r<2r_d$ the ratio has a large spread and its mean value
is larger than unity.  Because the average radial velocity of the gas
is close to zero, the large ratio means that there is substantial
pressure force.  The bottom panel shows the fractional difference
between the radial component of the total acceleration $g_t$ and the
gravitational acceleration $g$. The scatter shows that there are
hydrodynamical forces present.}  \label{fig:acceleration}
\end{figure}

We can get additional information about the status of the cold gas by
studying forces acting on individual gas particles. If the cold gas is
in a quiet rotation, the gravitation and the centrifugal forces should
be equal. (Note that the average radial velocity of the gas is close
to zero.)  The top panel in Figure~\ref{fig:acceleration} shows that
indeed at large radii ($r\simgreat 2r_d$) the average gravitational
acceleration of a particle is close to its centrifugal
acceleration. Yet, at smaller distances the situation is more
complex. On average the force of gravity is larger than $v^2_{\rm
rot}/r$. Because on average the gas does not move radially, this means
that there is another force, which keeps the particles from falling to
the center - the pressure force.   

The bottom panel in Figure~\ref{fig:acceleration} addresses the
issue of the non-gravitational forces in a more direct way: we plot
the deviations of the total acceleration $g_t$ from the gravitational
acceleration $g$. Again, on average $g_t\approx g$ at radii $r>r_d$
and there are asymmetries at smaller distances, where on average $g_t > g$.
What is quite remarkable about this plot is the magnitude of the
scatter. It is important to note that observed a factor of two
deviations in the acceleration do not translate to significant
deviations in velocities. For example, at $r=3r_d$ the rms velocity is
about 10~\kms\ -- only a 15\% deviation as compared with the rotation
velocity.  At the same distance the force deviations are a factor of
three. Here we are looking at the effect of the multiphase
medium. Particles of cold gas are pushed by the hot gas and get kicked
by supernovae explosions. Yet, particles do not move too far
(otherwise they would get large velocities). Before long, they get
pushed and kicked again in a different direction.

\end{document}